\documentclass[twocolumn,10pt]{aastex631}
\usepackage[utf8]{inputenc}
\usepackage{hyperref}
\usepackage{gensymb}
\usepackage{longtable}
\usepackage{siunitx}
\usepackage{graphics}
\usepackage{graphicx}
\usepackage{float}
\usepackage{ragged2e}
\usepackage{textcomp}
\usepackage{dirtytalk}
\usepackage{xcolor}
\usepackage{rotating}
\renewcommand{\arraystretch}{2}
\usepackage{fdsymbol}

\usepackage{lipsum}
\usepackage{tablefootnote} 

\shorttitle{Historical FAVA Flares}
\shortauthors{Joffre et al.}

\newcommand{\rad}{$R_{95\%}$}

\begin{document}

\title{Historical Fermi All-Sky Variability Analysis of Galactic Flares}


\author{S. Joffre} 
\affiliation{Department of Physics and Astronomy, Clemson University,  Kinard Lab of Physics, Clemson, SC 29634, USA}

\author{N. Torres-Alb\`a}
\affiliation{Department of Physics and Astronomy, Clemson University,  Kinard Lab of Physics, Clemson, SC 29634, USA}

\author{M. Ajello}
\affiliation{Department of Physics and Astronomy, Clemson University,  Kinard Lab of Physics, Clemson, SC 29634, USA}

\author{D. Kocevski}
\affiliation{ST12 Astrophysics Branch, NASA Marshall Space Flight Center, Huntsville, AL 35812, USA}

\author{R. Buehler}
\affiliation{Deutsches Elektronen Synchrotron DESY, D-15738 Zeuthen, Germany}

\begin{abstract}
    The \textit{Fermi} All-sky Variability Analysis (FAVA) provides a photometric alternative for identifying week-long gamma-ray flares across the entire sky while being independent of any diffuse Galactic or isotropic emission model. We reviewed 779 weeks of \textit{Fermi}-LAT data analyzed by FAVA to estimate the rate and origin of Galactic gamma-ray flares, and to search for new variable Galactic gamma-ray transients. We report an estimated yearly rate of $\sim8.5$ Galactic gamma-ray flares/year with $\sim1$ flare/year coming from unknown sources. Out of the known gamma-ray sources that are spatially coincident with these detected flares, we report gamma-ray flares for six of them for the first time. All six are classified as pulsars, or a source of unknown nature but which positionally overlaps with known supernova remnants or pulsar wind nebulae. This potentially means these sites are tentative candidates to be the second known site of a variable gamma-ray pulsar wind nebula (PWN), after the famous Crab Nebula's PWN. Additionally, we identify 9 unassociated flares that are unlikely to have originated from known gamma-ray sources.

\end{abstract}

\vspace{-3 cm}


\section{Introduction}

Launched in June of 2008, the \textit{Fermi Gamma-ray Space Telescope}  regularly surveys the gamma-ray sky with the Large Area Telescope (LAT, \citealp{atwood_fermilat}). \textit{Fermi}-LAT is able to observe the entire sky approximately every 3 hours. This semi-uniform exposure, as well as the LAT's stable instrument response, good angular resolution ($<0.15\degree$ at $>$10 GeV\footnote{Single photon, 68\% containment radius}), and large energy range (20 MeV $-$ 2 TeV) have made \textit{Fermi}-LAT an ideal instrument for studying time-varying phenomena in the gamma-ray sky. \par 

Ongoing observations by the LAT have shown that the GeV sky is populated by transient sources whose gamma-ray flux varies on an assortment of timescales. Gamma-ray bursts (GRBs) and pulsars' high-energy flux can vary in as short as a fraction of a second (\citealp{fermi_grb,fermi_2nd_pulsar}). Day-long flux variations have been observed in the Crab Nebula (\citealp{crab_abdo_2011,crab_balbo_2011,crab_striani_2011,crab_tavani_2011}) as well as in multiple novae (\citealp{Cheung_2016,Li_novae_2017,frankowiak_gray_novae,Nelson_novae_2019}). Blazars have been observed to emit gamma-rays whose emission timescales can vary from days and weeks (\citealp{2023ATel15925....1D,2023ATel15995....1C,2023ATel15952....1P,brill2023selfsupervised}) to months or even years (\citealp{1fgl,2fgl,3fgl,4fgl,Penil_2022,penil_2023}). 

Observations of the transient gamma-ray sky are done across multiple domains and collaborations. Searches for extremely short flares ($<1$ sec) like GRBs are performed onboard \textit{Fermi}, refined on ground, and made publicly available quickly through the Gamma-ray Coordinates Network (GCN) notices\footnote{ http://gcn.gsfc.nasa.gov/ipn.html}. Searches of transients on the 6-hour to 1 day timescales are performed manually on the ground by the \textit{Fermi} flare advocates and delivered to the community via Astronomer Telegrams (ATel) and refereed publications (\citealp{flare_advocates}). Sources whose variability timescale spans months-to-years are discovered and published in \textit{Fermi}-LAT catalogs (\citealp{1fgl,2fgl,3fgl,4fgl,4fgl-dr3}). As for flares with a typical duration of 1 week, these have been detected with the \textit{Fermi} All-sky Variability Analysis (FAVA; \citealp{1FAVA}) and are reported in the first and second catalog of Flaring Gamma-ray Sources from FAVA (1FAV \citealp{1FAVA}, 2FAV \citealp{2FAV}).



The vast majority of all the variable sources discovered by \textit{Fermi} are identified as blazars, a class of extremely variable active galactic nuclei (\citealp{urry_blazars_1998}). The very bright Galactic diffuse emission, which extends up to high Galactic latitudes and all along the Galactic plane, hampers the detection of Galactic transients. Ignorance of the sub-degree spatial structure of the Galactic diffuse emission coupled to systematic uncertainty in the LAT instrumental response (both the effective area and the point-spread function) makes source detection in the plane often difficult on all but the shortest ($< 1$s) timescales. This has strongly limited our knowledge of Galactic transients. Of the Galactic transients currently detected in gamma-rays (excluding pulsars and millisecond pulsars as they are a well-studied population, and their transient nature does not originate from flares), the following Galactic source populations are represented: gamma-ray binaries (\citealp{gray_binaries}), novae (\citealp{novae_as_gray_science}), and the Crab Nebula's Pulsar Wind Nebula (PWN; \citealp{grayflare_crab_ajello}). However, new detections even within these known source classes are rare. The Incremental \textit{Fermi} Large Area Telescope Fourth Source Catalog Data Release 4 (4FGL-DR4, \citealp{4fgl-DR4}), only reports the firm identifications\footnote{The 4FGL identification criterion depends on the source type. For pulsars or X-ray binaries they must find correlated periodic variability to be labeled as a `firm identification'} 
of 11 binary systems, 8 novae\footnote{Not all novae detected by \textit{Fermi}-LAT are reported in the 4FGL-DR4 due to the shorter duration of the flare. Koji Mukai's running list of all \textit{Fermi}-LAT detected novae can be found here: \url{https://asd.gsfc.nasa.gov/Koji.Mukai/novae/latnovae.html}. As of June 2023, 18 have been detected.}, and 13 PWN. If we include associations\footnote{A Bayesian (spatial coincidence only) and a Likelihood Ratio (spatial with log N-log S) method are used for associations. See Section 5 of \citealp{4fgl}}, the total number of sources represented by these three source classes is 60.  Galactic sources exhibiting transient behavior are even rarer. The Second Catalog of Flaring Gamma-ray Sources from the \textit{Fermi} All-sky Variability Analysis (2FAVA; \citealp{2FAV}) reports 3 high-mass binaries (HMBs), 5 novae, and 1 PWN (the Crab). The detection of a new Galactic transient has the potential to add significantly to the current data set due to the low statistics of observed Galactic transients.

In this work, we utilized FAVA to identify transients at low-latitudes ($|b|<10\degree$) that are likely to be of Galactic origin. We define the term `flare' to be a FAVA detected transient occurring over a week-long period with a positive flux variation. A `long-term flare' is defined to describe any set of spatially coincident FAVA flares that occur in consecutive weeks. We present our method, analysis, and results in the following manner: Section \ref{calibration} reviews how the positional 95\% uncertainty radius  (\rad) was calibrated while Section \ref{data} outlines how FAVA works and our selection criteria for flaring Galactic gamma-ray candidates. Following that, Section \ref{results} outlines the properties of the flares  coincident with Galactic gamma-ray sources, as well as with unassociated sources. Lastly, Section \ref{discussion} and Section \ref{conclusion} review and summarize this work.

\section{Uncertainty Radius Calibration}\label{calibration}

Currently, 1FAV and 2FAV include a $0.1\degree$ estimated systematic correction factor for a detected flare's uncertainty radius (\citealp{1FAVA,2FAV}). This systematic is based on the positions of known gamma-ray sources in comparison to the detected flares. Additional sources present in the 4FGL enable us to more accurately estimate the systematic uncertainty on a flare's positional. Following the method discussed at the end of Section 3.1.3 of the 4FGL paper (\citealp{4fgl}) we calibrated the maximum likelihood analysis method's derived \rad \, uncertainty regions for the FAVA detected gamma-ray flares. First, we cross-matched the FAVA flares that were spatially coincident with a known 4FGL gamma-ray source, giving us 7069 flares. This number reflects the removal of all FAVA detections that reported a \rad \, of 0.8$\degree$, which is the default value if the maximum likelihood analysis fails to converge properly. To estimate the systematic uncertainty we introduce a relative systematic factor $f_{rel}$ such that 
\begin{equation}
    R_{95 tot} = f_{rel}R_{95 stat}
\end{equation}

where $R_{95tot}$ is the total error (at the 95\% confidence level) and $R_{95stat}$ is the uncertainty error in the position.

\begin{figure}{h}
    \centering
    \includegraphics[width=0.4\textwidth]{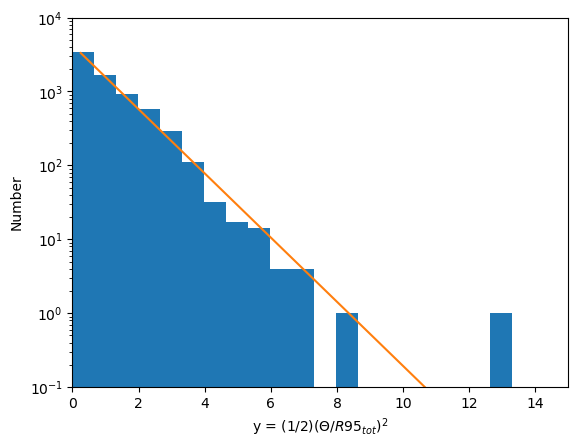}
    \caption{Best fit for calibration of \rad \, values with Eqn. \ref{rayleigh} plotted in log-log space. $f_{rel}= 1.06$ }
    \label{fig:calibration}
\end{figure}

Defining $\Theta$ as the angular separation between the FAVA detected source location and the counterpart location of the source, we test various values of $f_{rel}$ until the value $y$ as defined by

\begin{equation}\label{rayleigh}
    y = \frac{1}{2} \left(\frac{\Theta}{R_{95tot}}\right)^2
\end{equation}

follows a Rayleigh distribution (i.e. $exp(-y)$). 
Our best fit is achieved for $f_{rel}=1.06$ and is plotted in Figure \ref{fig:calibration}. Therefore, all reported \rad\, have been increased by 6\% over their original values reported by FAVA.

\section{Data Selection}\label{data}
All data was procured from the \textit{Fermi} All-sky Variability Analysis (FAVA) website\footnote{\url{https://fermi.gsfc.nasa.gov/ssc/data/access/lat/FAVA/}}. The last week of data reported in this paper is week 780  (July 17th, 2023). 

\subsection{FAVA}

The \textit{Fermi} All-sky Variability Analysis (FAVA; \citealp{1FAVA}) identifies transients across the entire sky using a photometric analysis method. 
In this method, FAVA does not rely on any Galactic diffuse model. Instead, FAVA assumes that the diffuse gamma-ray emission is constant over long time periods ($>$ weeks). FAVA searches for transients by comparing measured counts over a week-long time bin at every point in the sky to the average emission over the first 4 years of the \textit{Fermi} mission. FAVA does so while accounting for the difference in exposure time of the sky, and the point spread function (PSF) dependence on both energy and the instrument off-axis angle. Observed counts above this 4 year averaged emission are analyzed and are converted into a probability of a flare being present. All analysis on the week-long time bins is broken into two energy bands: a low energy band (LE; 100$-$800 MeV) and a high energy band (HE; 800$-$300,000 MeV). \citealp{2FAV} reports a photometric LE flux sensitivity of $F_{LE} = 3.67 \times 10^{-7}\text{cm}^{-2}\text{s}^{-1}$ and a HE flux sensitivity of $F_{LE} = 3.24 \times 10^{-8}\text{cm}^{-2}\text{s}^{-1}$. To better localize the origin of the flare, a maximum likelihood analysis is performed at its position. For an in-depth explanation of this and FAVA please see \citealp{2FAV}.  \par


\begin{deluxetable*}{|c| c| c| c| }[h]
\centering
   \tablecaption{Cuts On FAVA Dataset For Determining Unassociated Flares
    }\label{tab:fava_cuts}
    \tablehead{\colhead{Cut Number} & \colhead{Cut Criterion}  & \colhead{Number of Flares Cut} &\colhead{Number of Unassoc. Flares Remaining}
    }   
\startdata
\hline
1 & FAVA Week 1$-$780 & 0 & 24832  \\
 \hline
2 & $|b|<10\degree$ & 22622& 2210  \\
 \hline
3 & LE or HE Fava detection $>5\sigma$ & 1380 & 830  \\
 \hline
4 & Unassoc, Unassoc. and Gal. or no 4FGL coincidence & 687 & 143  \\
 \hline
5 & Coincident with 2FAV blazar & 12 & 131  \\
 \hline
6 & WISE IR blazar colors & 120 & 11 \\
\hline
7 & Solar Vicinity & 0 & 11  \\
\hline
8 & CHIME, BAT, BGM: Gamma-ray burst cross-match &1 &10  \\
\hline
\enddata
\footnotesize
\textbf{Table \ref{tab:fava_cuts}}\\
See Section \ref{pipeline_filter} for a detailed description. For Cut 4, 105 flares are exclusively coincident with a known Galactic 4FGL-DR4 source or a gamma-ray emitting nova. Ten additional flares are spatially coincident with 1 or more Galactic 4FGL-DR4 source, or a gamma-ray emitting nova. All of these flares have a solar distance $>16\degree$. 
\end{deluxetable*}
\vspace{-2.5cm}
\subsection{Filter for Identifying Galactic and Unassociated Flare Candidates}\label{pipeline_filter}
To minimize contamination in our data set from extragalactic sources the following cuts were made. The number of detected flares remaining after the cut is reported. Table \ref{tab:fava_cuts} is a summary of the cuts utilized to identify the number of unassociated flaring Galactic gamma-ray candidates. We note that the \rad \, used for the FAVA detected flare is from the localized maximum likelihood analysis of the flare where the HE or LE band is used depending on which one is better localized. We have calibrated these \rad \, as discussed in Section \ref{calibration}. 
\begin{enumerate}
    \item All detected flares  from the FAVA website from week 1 to week 780\footnote{Week 768 is excluded due to analysis complications from the FAVA pipeline so the data is not accessible.} (August 4th, 2008 $-$ July 17th, 2023) at a threshold of $>5$\footnote{This is not equivalent to a 5$\sigma$ Gaussian significance. See the github page for notes} are downloaded\footnote{See \url{https://github.com/dankocevski/pyFAVA} and  Appendix for more details} $-$ 24,832 flares
    
    \item All flares with a Galactic latitude $|b|<10\degree$ are retained $-$ 2,210 flares remain
    
    \item Only flares whose low-energy and/or high-energy FAVA detection is at least 5$\sigma$ are saved $-$ 830 flares remain

    \item The remaining flares are then positionally cross-matched with the 4FGL-DR4 (\citealp{4fgl-DR4}), Roma-BZCAT (\citealp{5bzcat}) as well as the list of definitive gamma-ray emitting novae by Koji Mukai\footnote{Included in the downloadable files for this paper.}. We break this down into 4 categories:
        \begin{itemize}
            \item Flares positionally coincident with a 4FGL-DR4 blazar or a blazar reported in the Roma-BZCAT. These flares are removed.
            \item Flares exclusively  coincident with the position of a known 4FGL-DR4 Galactic source or a gamma-ray emitting novae (i.e. are not spatially coincident with an unassociated or known blazar source). These flares are set aside and are categorized as originating from known Galactic sources.  
            \item Flares positionally coincident with a 4FGL-DR4 unassociated source, or with a 4FGL-DR4 unassociated source and a known Galactic source 
            \item Flares that do not positionally coincide with a gamma-ray source reported in the 4FGL-DR4. 
        \end{itemize}

    The 4FGL-DR4 does not provide gamma-ray \rad \, for some of the detected novae, as well as not detecting others, hence, why we utilize the compiled gamma-ray novae list. This second subset (flares exclusively coincident with the position of a known 4FGL-DR4 Galactic source or a gamma-ray emitting novae) is composed of 105 individually unique FAVA flares. This subset is set aside and will be discussed at the end of this section. The last two items encompass possibly new, or unknown sources, and are the subset of flares we discuss for the remainder of the data selection cuts. $-$ 143 flares
    
    \item Flares that are coincident with a 2FAV source that is associated with a known blazar are eliminated $-$ 131 flares remain.
    
    \item Following the Wide-Field Infrared Survey Explorer (WISE) blazar strip as reported in \citealp{Massaro_2012}, we developed a code\footnote{See \url{https://github.com/tryingastronomy/Blazar_codes/tree/main/AW_blazar_strip} and Appendix for more details} to filter out any flare detection whose \rad \,\, contains a WISE source (from the AllWISE catalog) with infrared colors within the blazar strip in all 3 dimensions (W1$-$W2 vs. W2$-$W3, W2$-$W3 vs. W3$-$W4, and W1$-$W2 vs. W3$-$W4) as this is considered a likely blazar. Upper limits on reported AllWISE magnitudes are considered. $-$ 11 flares remain.

    \item FAVA reports the real-time angular distance of the flare to the position of the Sun at the time the flare was detected. Following \citealp{1FAVA}, since the Sun is a bright gamma-ray emitting source, we cut all flares whose Sun distance is less than 12$\degree$. None of these remaining flares met this criteria so none were cut $-$ 11 flares remain. 


    \item Lastly, in order to see if the unassociated FAVA detected flares are coincident with other transient detections we cross-matched our remaining flares with the First Chime/FRB Fast Radio Burst catalog (\citealp{chime_frb_cat1}), the \textit{Swift}/BAT Hard X-Ray Transient Monitor (\citealp{swiftbattransient}), the Third \textit{Swift} Burst Alert Telescope Gamma-Ray Burst (GRB) Catalog  (\citealp{swiftbat_grb}), the Fourth Fermi-GBM Gamma-Ray Burst Catalog (\citealp{von_Kienlin_2020}) and the \textit{Fermi}-LAT catalog of long-term gamma-ray transient sources (1FLT; \citealp{1FLT_catalog}). The only spatial cross-matches that occur is with the Third \textit{Swift} Burst Alert Telescope (BAT) Gamma-Ray Burst (GRB) Catalog, and the Fourth Fermi-GBM Gamma-Ray Burst Catalog. Both catalogs report the detection of GRB 221009A (RA: 288.26, DEC: 19.77; 2022 October 9 at 13:16:58.99 UTC (\citealp{BOAT_GRB})) which is both spatially and temporally coincident with one of the 11 remaining unassociated FAVA flares (flare 7401). This GRB is the recently detected brightest-of-all-time (BOAT) GRB with the highest total isotropic-equivalent energy and highest fluence and peak flux GRB ever identified (\citealp{burns_boat}). Therefore, this flare is removed $-$ 10 flares remain.

    \item Flares that have coincident positions based on overlapping \rad \,  are assumed to be flaring events likely originating from the same source. Only two of the 10 unassociated flares spatially overlap. See Table \ref{tab:unassoc} $-$ We report 9 unique, new flaring objects of unknown origin.

\end{enumerate}
Over almost 15 years of data, we find a total of 9 distinctly new flaring sources (composed of 10 flares) that have not been associated to any known gamma-ray detected source and are of potential Galactic origin (see Table \ref{tab:unassoc}). We find an additional 105 flares that are coincident with a single Galactic gamma-ray source, plus 10 flares that are coincident with more than one, resulting in a total of 115 flares coincident with at least 1 Galactic gamma-ray source (see Table \ref{tab:galdr4_src_fav_det}). This means we have 125 flares of likely Galactic origin. 
Section \ref{flare_estimates} discusses the yearly flare rate more thoroughly.   \par

The flares that are spatially associated with Galactic gamma-ray sources, novae, and unassociated flares are all outlined in three separate tables. Table \ref{tab:galdr4_src_fav_det} reports on non-novae concomitant with FAVA detections, Table \ref{tab:gray_novae} reports on  the FAVA detections of classical novae,  and Table \ref{tab:unassoc} reports on FAVA flares that we propose as originating from unidentified Galactic transient sources.  \par

 
\subsection{Rate of Spurious FAVA Detections}
Since we are selecting flares on a 5$\sigma$ statistical significance, this leaves a margin for spurious detections. To evaluate the likelihood of one of our FAVA detected flares being spurious, we first calculate the total number of trials. The number of trials can be calculated using
\begin{equation}
    N_{trials} = N_{weeks} \times N_{channels} \times N_{sky\, pixels}
\end{equation}
where $N_{trials}, N_{weeks}, N_{channels}, \text{and } N_{sky \, pixels}$ all correspond to the number of trials, weeks, energy channels, and sky pixels used in the analysis, respectively. With $N_{weeks}=779, N_{channels} = 2,$ and $N_{sky \, pixels} = 165,012$ where 
\begin{equation}
    N_{sky \, pixels } = \frac{\text{Total Sky}}{\text{Pixel}} = \frac{41,253\deg^2}{0.25\deg^2} = 165,012
\end{equation}
resulting in $N_{trials}=257,088,696$. Since a $5\sigma$ detection corresponds to a probability $P=3\times10^{-7}$, this means there is a 1/3,333,333 chance of a wrongful detection. Dividing our number of trials by 3,333,333 we expect 77 spurious FAVA detections out of 257,088,696 trials. From our 779 weeks analyzed, we find 13,218 flares were detected at a $5\sigma$ level, meaning we expect 0.5\% of all FAVA detections to be false. Hence, we do not expect any of our vetted flares to be of spurious origin.

\subsection{Background Extragalactic Contamination Estimates}
\citealp{1FAVA} found that by simulating extragalactic sources at low latitudes (after accounting for differences in solid angle and sensitivity corrections), 60\% of low-latitude, variable extragalactic sources would be able to be detected by LAT. The 2FAV also reports a total of 323 flares at $|b|<10\degree$, of which, 249 are associated to an extragalactic source. We therefore anticipate that of the 2,210 flares our investigation finds at $|b|<10\degree$, 1701 are likely of extragalactic origin. After all of our cuts that were implemented to eliminate extragalactic contamination, we removed over 2,000 flares (with 125 Galactic gamma-ray flare candidates remaining). The number of removed flares means it is very likely that we have properly accounted for this potential contamination on a statistical basis.

\subsection{Robustness of the 3D WISE blazar Strip}\label{pipeline robustness}


Filter criterion No. 6 in Section \ref{pipeline_filter}, the 3D WISE Blazar strip (W1$-$W2, W2$-$W3, W3$-$W4; \citealp{Massaro_2012}), was tested using 1) 4FGL-DR4 blazars 2) 4FGL-DR4 Galactic sources, and 3) known \textit{Fermi}-LAT detected gamma-ray novae 

\subsubsection{Testing Known Blazar Sources}\label{sec:test_blazars}
The 4FGL-DR4 contains 3,934 blazars of the BLL, FSRQ, and undetermined type. In this catalog, the gamma-ray detection is reported as well as the associated counterpart, if known. Using the \rad\, of the 4FGL-DR4 gamma-ray detection, we verify whether or not a WISE source contained therein  has WISE colors that would place it within all three dimensions of the blazar strip. Of the 3,934 known blazars in the 4FGL-DR4, $\sim 96$\% 
are compatible (considering the upper limits) with the blazar strip. \par

Of the known blazars, $\sim 89\%$ are at high latitude, ($|b|>10\degree$). This leaves 420 known blazars at a latitude of $|b|<10\degree$. Of the blazars that lie in the plane, 
$\sim$76\% are captured in all three dimensions of the blazar strip. \par 

\subsubsection{Testing Known Galactic Sources as reported by 4FGL-DR4}\label{sec:test_gal}


The 4FGL-DR4 reports 584 sources of Galactic origin (pulsars, supernova remnants (SNR), high mass binaries (HMBs), etc.). Of the reported Galactic sources, 
405 occur at low latitudes ($|b|<10\degree$). 
Of the known Galactic sources at low latitude (as some are at $|b|>10\degree$), $\sim 39\%$  have uncertainty regions which contain a WISE source whose colors are consistent with a blazar. This means our method wrongly captures  $\sim 39\%$ of low-latitude Galactic sources as blazars.  Many of these WISE sources report upper limits (i.e. the value of the magnitude is a minimum) for the different WISE filters. If one of the WISE source's color values has the possibility to fall in the blazar strip because of this limit, we count it as a possible blazar. 

\subsubsection{\textit{Fermi}-LAT Detected Gamma-ray Novae }
We also test Koji Mukai's compiled list of \textit{Fermi}-LAT detected classical novae mentioned in the introduction. Of the 18 definitive detections (reported as of October 10th, 2023), only 10 have a WISE source within 5 arcseconds of the optically detected location. Out of these 10 sources, only 1 falls in all three blazar strips (V407 Cyg, which we detect with FAVA. See Table \ref{tab:gray_novae}). Since we assume that the lack of a WISE source means that the source in question is not a blazar, then we only mistake gamma-ray novae for blazars 6\% of the time. 



\subsubsection{Sensitivity and Specificity of 3D WISE blazar Strip}\label{sensitivity_specificity}

Since we are particularly interested in this test working at low latitudes ($|b|<10\degree$) we report the results of low latitude sources reported in the 4FGL-DR4. For clarification: 

\begin{itemize}
    \item True positives (TP) are low latitude blazars flagged as blazars
    \item False positives (FP) are low latitude Galactic sources flagged as blazars
    \item True negatives (TN) are low latitude Galactic sources which have no WISE sources in their \rad \, with colors consistent with blazars in all three dimensions of the WISE blazar strip. Therefore these are not flagged by the blazar strip (because they are not blazars).
    \item False negatives (FN) are low latitude blazars whose \rad \, contains no WISE source with colors consistent with a blazar in all three dimensions of the WISE blazar strip. Therefore this blazar is not flagged by the blazar strip as a blazar.
\end{itemize}

Sensitivity, or true positive rate (TPR) is defined as
\begin{equation}
    \text{TPR} = \frac{\text{TP}}{\text{TP + FN}}
\end{equation}

while specificity, or true negative rate (TNR) is defined by
\begin{equation}
    \text{TNR} = \frac{\text{TN}}{\text{TN + FP}}
\end{equation}

For our purposes, sensitivity reflects the ability of the test to correctly identify a source as a blazar. Specificity, on the other hand, reflects the ability of the test to identify non-blazar sources (i.e. Galactic sources). 
From the tests in Section \ref{sec:test_blazars} and Section \ref{sec:test_gal} we find the following values

\begin{itemize}
    \item TP $ = 318$ blazars flagged as blazars
    \item FP $= 159$ Galactic sources flagged as blazars 
    \item TN $= 246$ Galactic sources not flagged as blazars 
    \item FN $= 102$ blazars not flagged as blazars 
\end{itemize}
which results in a sensitivity (or TPR) of $75.7\%$ and a specificity (TNR) of 
60.7\% . These values can be combined to determine the test accuracy (ACC) 
\begin{equation}
    \text{ACC} = \frac{\text{TP + TN}}{\text{TP + TN + FP + FN}}
    \label{acc}
\end{equation}

which, when the above values are evaluated in equation \ref{acc} the result is ACC$=68.4\%$. The accuracy value reflects the fraction of blazars and Galactic sources which are properly flagged compared to the entire set that is analyzed.

\section{Results}\label{results}

\subsection{Overview of FAVA Flare Properties Coincident with Galactic Sources} \label{sec:overviewFavGal}

We find 115 FAVA-detected flares that are spatially coincident with at least 1 4FGL Galactic gamma-ray source, with 105 flares exclusively coincident (spatially) with a single Galactic source. We have determined that all 115 flares spatially coincide with only  19 known Galactic gamma-ray sources. Approximately $\sim75\%$ of the 105 flares that are spatially coincident with a single Galactic gamma-ray source can be attributed to Cygnus X-3, the Crab Nebula, and the high-mass binary (HMB) system of PSR B1259-63. See Section \ref{sec:gal_assoc} for more discussion. Overall, the source types that are spatially coincident with a FAVA flare are: high-mass binary (HMB), Pulsar wind nebula (PWN), pulsar (PSR), nova (NOV), millisecond pulsar (msp), and potential association with supernova remnant or pulsar wind nebula (spp). \par

The sources reported in Table \ref{tab:galdr4_src_fav_det} are typically best fit with a \texttt{LogParabola} model in the 4FGL-DR4, but for the sake of comparison with the FAVA fits, we only report the power law fit for each source from the 4FGL. Additionally, we report the averaged FAVA LE and FAVA HE band indices from the FAVA detected flares that are spatially coincident with the 4FGL source. Out of the known Galactic sources detected by FAVA, the hardest spectrum reported by the DR4 is PSR B1259-63 with $\Gamma_{4FGL} = -2.33 \pm 0.27$. The hardest LE band source is the PSR J1826-1256 or SNR G018.1-00.1 at $-2.14\pm0.28$. 
Of flares spatially coinciding with a 4FGL Galactic source, the only flare detected at a $>5\sigma$ level in the HE band is PSR B1259-63. 


Among the 18 known gamma-ray novae so far, using our selection criteria, 8 were detected by FAVA (see Table \ref{tab:gray_novae}). These are V407 Cyg, V959 Mon, V1324 Sco, V339 Del, V5855 Sgr, V5856 Sgr, V906 Car, and V392 Per. Two of these novae had FAVA flares that were coincident with other sources. Flare Identification number (FID) 862, which we have identified as V407 Cyg, is also positionally coincident with SNR G085.9$-$00.6. But, due to the temporal coincidence during the height of the gamma-ray emission from V407 Cyg (see \citealp{v407cyg_2010gray}), this flare most likely originated from the nova. Similarly, FID 4311, encompasses the positions of both V5855 Sgr and the blazar of uncertain type (bcu) NVSS J181120-275946. The \texttt{Variability\_Index} (as reported by the 4FGL-DR4) of this blazar is 16.27, which falls below the threshold (27.69) at which the source has a $99\%$ chance of being variable. This flare is detected in the same week that V5855 Sgr peaked in gamma-rays (\citealp{2016ATel.9736_v5856sgr,2016ATel.9699_v5855sgr,V5855Sgr_Munari_2017}). We therefore conclude that this flare originated from V5855 Sgr. \par


Selecting the most significant detections from the novae with multiple detections, 6 novae were detected above $5\sigma$ in both LE and HE bins while the remaining 3 were detected only in the HE bin (see Table \ref{tab:gray_novae}). No gamma-ray detected nova flares were detected exclusively in the LE bin. The average LE photon index of all of these novae is $\Gamma_{LE} = -1.49\pm0.39$, while the average HE photon index is $\Gamma_{HE} = -2.57\pm0.36$. The stark difference in slope between the LE and HE bin (being hard and soft, respectively) points to these novae emission peaking between the two bins. Since the dividing line is at 800 MeV, we expect that many of these gamma-ray detected novae are displaying a cutoff around a few GeV, which has been reported in previous analyses (\citealp{fermi_lat_novae,nova_shock_cutoff}).

%





\begin{longrotatetable}
\begin{deluxetable*}{c c c c c c c c c c }
\centering
   \tablecaption{DR4 Galactic Sources Coincident with FAVA Galactic Flare Candidates$^\dag$
    }\label{tab:galdr4_src_fav_det}
    \tablehead{\colhead{Associated} &  \colhead{4FGL Name} & \colhead{DR4 Source} & \colhead{LE $\sigma^*$} & \colhead{HE $\sigma^*$} & \colhead{Exclusive} & \colhead{Max Possible} &  \colhead{$\Gamma_{4FGL}^{d}$} & \colhead{$\Gamma_{LE_{avg}}^{f}$} & \colhead{$\Gamma_{HE_{avg}}^{f}$} \\
    \colhead{Source} &  \colhead{} & \colhead{Class$^{a}$} & \colhead{} & \colhead{} & \colhead{Flare Number$^{b}$} & \colhead{Flare Number$^{c}$}    
    }   
\startdata
\hline
Cygnus X-3 & 4FGL J2032.4+4056 & HMB & 11.89 & 4.78 & 30 & 36  & $-2.68 \pm 0.04^{\clubsuit}$ & $-2.52\pm0.22$ & $-2.56\pm0.52$\\ 
Crab Nebula (IC Field) & 4FGL J0534.5+2201 & PWN  & 23.56 & 4.29 & 28 & ... & $N/A^{\clubsuit}$$^{\triangle}$ & $-3.38\pm0.34$ & $-2.26 \pm 0.73$  \\ 
PSR B1259$-$63&4FGL  J1302.9-6349& HMB & 18.41  & 6.84  & 21 & ... & $-2.33\pm0.27$ & $-2.83\pm0.20$ & $-2.17\pm0.69$ \\ 
LSI +61 303 &  4FGL J0240.5+6113 & HMB & 7.08 & 2.49  & 7 & 8  & $-2.39\pm0.005^{\clubsuit}$ & $-2.33\pm0.27$ & $-1.85\pm0.64$ \\ 
PSR J0248+6021 & 4FGL J0248.4+6021 & PSR & 6.29  & 1.07  & 3 & 4 & $-2.64\pm0.02^{\heartsuit}$ & $-2.96\pm0.33$& $-3.01\pm0.59$ \\ 
PSR J2032+4127 & 4FGL 2032.2+4127 &PSR & 8.51 & 3.97  & 2 & 8 & $-2.61\pm0.01^{\heartsuit}$ & $-2.61\pm0.19$ & $-2.73\pm0.56$ \\ 
PSR J1731-1847 & 4FGL J1731.7-1850 & msp & 5.01  & -0.46  & 1 & ... &$-2.65 \pm 0.09^{\clubsuit}$& DNC$^{\times}$ & DNC$^{\times}$\\ 
\hline
SNR G016.0$-$00.5$^{\dag}$& 4FGL J1821.4-1516  & spp  & 5.55 & 4.03 & 1 & ...&$-2.90\pm0.09^{\clubsuit}$ & $-2.96\pm0.32$ & $-0.06 \pm 0.78$ \\ 
SNR G016.7+00.1$^{\dag}$& 4FGL J1821.1-1422 & spp  & 5.55 &  4.03 & ... & ... &$-2.66\pm 0.06^{\clubsuit}$ & ... & ...  \\ 
\hline
PSR J1826-1256 $^{\ddag}$ & 4FGL J1826.1-1256 & PSR & 9.66 & 1.04  & 1 & ... & $-2.46\pm0.01^{\heartsuit}$ & $-2.14\pm0.28$ & $-1.59\pm0.54$\\ 
SNR G018.1-00.1 $^{\ddag}$ &  4FGL J1824.1-1304 & spp & 9.66 &  1.04 & ... & ... &$-2.94\pm0.09^{\clubsuit}$ & ... & ... \\ 
\hline
\enddata
\vspace{0.05cm}
\footnotesize
\textbf{Table \ref{tab:galdr4_src_fav_det}} \\
$^{a}$ Capitalized classifications are confirmed associations, lowercase are positionally coincident associations, high-mass binary (HMB), Pulsar wind nebula (PWN), pulsar (PSR), Nova (NOV), millisecond pulsar (msp), potential association with supernova remnant or pulsar wind nebula (spp) \\
$^{b}$ - Number of FAVA flares that are spatially coincident with only this Galactic source and no others.  \\
$^{c}$ - Total possible number of FAVA flares that could have originated from this Galactic source, where we include flares that are coincident with this and at least 1 other DR4 known Galactic source. `...' corresponds to the maximum number being the same as the Exclusively Coincident Flare Number \\
$^{d}$ - 4FGL-DR4 reported power law photon index \\
$^{f}$ -  Average photon index of all  FAVA flares exclusively coincident with the source listed.\\ 
$^{\dag}$ - These Galactic sources are spatially coincident with the same flare, FID 3472  \\
$^{\ddag}$ - These Galactic sources are spatially coincident with the same flare, FID 3473. Two additional unassociated 4FGL sources are also coincident with the edge of the flare: 4FGL J1826.5-1202c and 4FGL J1828.1-1312 \\
$^{\clubsuit}$ - 4FGL-DR4 best fit as LogParabola \\
$^{\triangle}$ - 4FGL-DR4 does not report a power law fit for the inverse Compton emission of the Crab PWN. It is fit with a LogParabola model with \texttt{LP\_Index} = 1.75 and \texttt{LP\_beta} = 0.08. \\
$^{\heartsuit}$ - 4FGL-DR4 best fit as PLSuperExpCutoff \\
$^*$ - Highest detection recorded. Note, may not be the same flare for LE $\sigma$ and HE $\sigma$ column \\
$^\times$ - Follow-up maximum likelihood analysis did not converge (DNC)\\
\end{deluxetable*}
\end{longrotatetable}











\begin{figure*}
    \centering
    \includegraphics[width=1.08\textwidth]{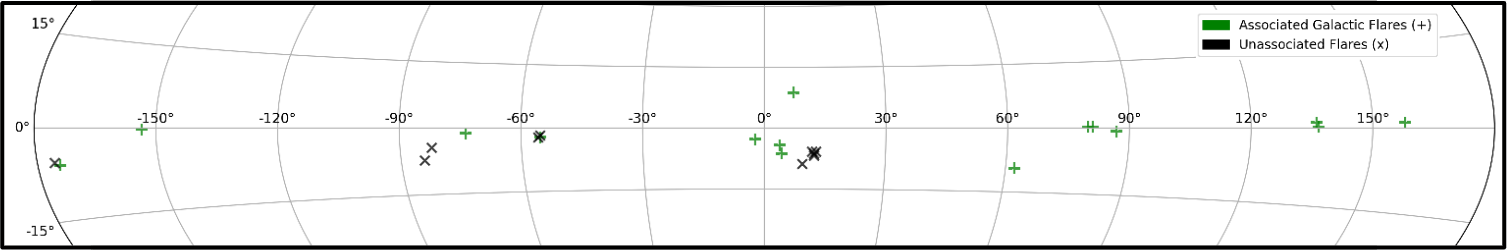}
    \caption{Sky distribution of detected flares. A single exclusively coincident FAVA flare that is spatially associated with a Galactic gamma-ray source or nova from Table \ref{tab:galdr4_src_fav_det} and Table \ref{tab:gray_novae} is represented by one of the 15 green crosses (as 4 sources are non-exclusive with the spatially coincident flare their flares are not included). The 10 black `X's correspond to the 10 unassociated Galactic flare candidates in Table \ref{tab:unassoc}. } 
    \label{fig:skydist}
\end{figure*}

\begin{longrotatetable} 
\begin{deluxetable*}{l l l c c c c c c c c}
\centering
   \tablecaption{Gamma-ray Novae Coincident with FAVA Galactic Flare Candidates
    }\label{tab:gray_novae}
    \tablehead{\colhead{Name}  & \colhead{Date of $\gamma$-ray Peak$^a$} &\colhead{Paper/ATel} & \colhead{Ang. Sep$^b$} &  \colhead{FAVA \rad$^c$} & \colhead{FAVA Dates of Detection} & \colhead{FIDs$^d$} & \colhead{LE $\sigma$} & \colhead{HE $\sigma$}  & \colhead{$\Gamma_{LE}$} & \colhead{$\Gamma_{HE}$} \\
    \colhead{} & \colhead{Y/M/D} & \colhead{} & \colhead{($\degree$)} & \colhead{(\degree)}  & \colhead{Y/M/D}  
    }   
\startdata
\hline
V407 Cyg & 2010/3/13-14 & \citealp{v407cyg_2010gray}& 0.12 & 0.08 & 2010/3/11-22 & \textbf{843},853,862 & 5.64$^*$ & 10.27$^*$ & $-1.37\pm0.25^*$ & $-2.52\pm0.20$$^*$  \\
V959 Mon & 2012/06/22 - 6/24 & \citealp{2012ATel.4224_v959mon} & 0.06 & 0.16 & 2012/6/18-25 & 2033&  7.39& 8.27 & $-1.82\pm 0.23$ & $-3.33\pm0.43$ \\
V1324 Sco & 2012/6/15 - 7/2 & \citealp{2012ATel.4284_v1324} & 0.06 & 0.07 & 2012/6/18-25  & 2034 & 4.15& 6.43 & $-1.17\pm0.33$ & $-2.54\pm0.24$ \\
V339 Del & 2013/08/16 & \citealp{2013ATel.5302_v339del} & 0.07 & 0.08 & 2013/8/12-19 & 26314, \textbf{26423} & 8.73$^*$ & 10.76$^*$ & $-1.25\pm0.22$$^*$ & $-2.54\pm0.24$$^*$ \\
V5855 Sgr & 2016/10/28-11/1 & \citealp{2016ATel.9699_v5855sgr} & 0.31 & 0.80 & 2016/10/31 - 11/7 & 4311 & 4.71 & 5.22 & DNC & DNC \\
V5856 Sgr & 2016/11/8 & \citealp{2016ATel.9736_v5856sgr} & 0.22 & 0.80 & 2016/11/7-14 & 43231 & 6.82 & 14.64 & DNC & DNC \\
V906 Car & 2018/4/10-4/14  & \citealp{aydi2020direct} & 0.09 & 0.04 & 2018/4/16-30  & \textbf{5071},5082 & 19.05$^*$ & 24.76$^*$ & $-1.81 \pm 0.09$$^*$ & $-2.62 \pm 0.09$$^*$ \\
V392 Per & 2018/4/30-5/8  & \citealp{v392per_2022} & 0.04 & 0.07  & 2018/4/20 - 5/7 & 5091 & 4.55 & 6.38 & $-0.55\pm0.76$ & $-2.19\pm 0.22$\\
\hline
\enddata
\vspace{0.05cm}
\footnotesize
\textbf{Table \ref{tab:gray_novae}} \\
$^a$ Date as given by corresponding paper or ATel (Astronomer's Telegram)\\
$^b$ Average angular separation between the optically detected Novae location and the FAVA detected flare. Rounded up to hundreths of degree\\
$^c$ If multiple flares, the best (smallest) localized \rad \, is reported\\
$^d$ Bolded text indicates the FID corresponding to the most significant detection (in both bands)\\
$^*$ Corresponds to the flare of highest significance
\end{deluxetable*}
\end{longrotatetable}

\subsection{Associated Galactic Sources Detected by FAVA With Previous Flaring Behavior}\label{sec:gal_assoc}
In the following subsections, we outline our findings of flaring behavior for the number of sources previously known to vary that are spatially coincident with FAVA flares. Furthermore, we outline the previously understood flaring nature of these sources. We reiterate, the term `flare' is meant to be a FAVA detected flare occurring over a week-long period unless otherwise noted. A `long-term' flare is meant to describe any set of spatially coincident FAVA flares that occur in consecutive weeks.   

\subsubsection{Cygnus X-3}\label{sec:cygx-3}
In this investigation, we report 30 FAVA flares exclusively coincident (spatially) with this source, and 6 additional flares that are also coincident with PSR J2032+4127. In this investigation we find Cygnus X-3 to have the highest number of coincident FAVA detected flares out of any known Galactic gamma-ray source. Of these 30 week-long flares, 10 occur in 2020 while 9 occur in 2021, with a total of 7 long-term flares. \par

Cygnus X-3 is a high-mass binary and the first  microquasar detected by \textit{Fermi}-LAT (\citealp{cygnus_x3}). Composed of a neutron star or black hole (\citealp{Koljonen_2017}), this is one of the few binaries of this type where the donor star is a Wolf-Rayet star (\citealp{cygx3_wolf_rayet}). Furthermore, \citealp{cygnus_x3} found Cygnus X-3 to have a  4.8 hour orbital period. Recent work done by \citealp{Prokhorov_2022_newestongaltransients} used a variable-size sliding-time-window (VSSTW) analysis to detect 23 intervals of 28-day windows where Cygnus X-3's gamma-ray emission was detected at levels $>4\sigma$.  \par

In 2021, 4 FAVA detected flares occur consecutively, from 2021 May 17 to 2021 June 14, likely detecting a single long-term flare.  We find that in 2020, the 10 flares do not happen all consecutively, with the longest long-term flare happening from 2020 August 24 to 2020 September 14  (3 weeks). \citealp{Prokhorov_2022_newestongaltransients} also reports a single long-term flare between 2020 April 27 and 2020 September 14.  Lastly, \citealp{Prokhorov_2022_newestongaltransients} reports the detection of a long-term flare from Cygnus X-3 during  14 September 2020 to 12 August 2021. FAVA detects 11 (non-consecutive) flares during this time.  



\subsubsection{Crab Nebula (PWN)}
Coming in second for the most FAVA flares spatially coincident with a Galactic gamma-ray source, this investigation reports 28 spatially coincident FAVA flares with the Crab Nebula. Long-term flares are detected three times. \par

One of the most well-studied gamma-ray sources on the sky, the Crab, is composed of a pulsar at its center surrounded by a nebula and the scattered remains of the supernova that occurred in 1054 A.D. (\citealp{1942ApJ....96..199M}). Even before the first FAVA catalog paper (\citealp{1FAVA}), gamma-ray flares from the Crab Nebula had been observed (\citealp{grayflare_crab_ajello,crab_tavani_2011,buehler_crab_flare_marco}). \citealp{grayflare_crab_ajello} identified a 16 day flare and 4 day flare in February 2009 and September 2010, respectively. Later, \citealp{crab_waves_2013} proposed the transient gamma-ray nature of the Crab to come from  flares (above-average flux on the order of hours) and `waves' (above-average outbursts occurring on the order of 1-2 weeks), the latter being formed by plasma instabilities with enhanced magnetic fields. \citealp{crab_waves_2013} records 7 different wave events from \textit{AGILE} and \textit{Fermi}-LAT data, 1 of which corresponds to 1 of our 28 FAVA flares that are exclusively coincident with the Crab. Labeled $W6$ in \citealp{crab_waves_2013}, this wave-natured flare is reported to have lasted 12 days from 2012 March 02 to 2012 March 14. The temporally coincident FAVA detection occurred in the week from 2012 March 05 to 2012 March 12.

\begin{figure*}
    \centering
    \includegraphics[width=0.8\textwidth]{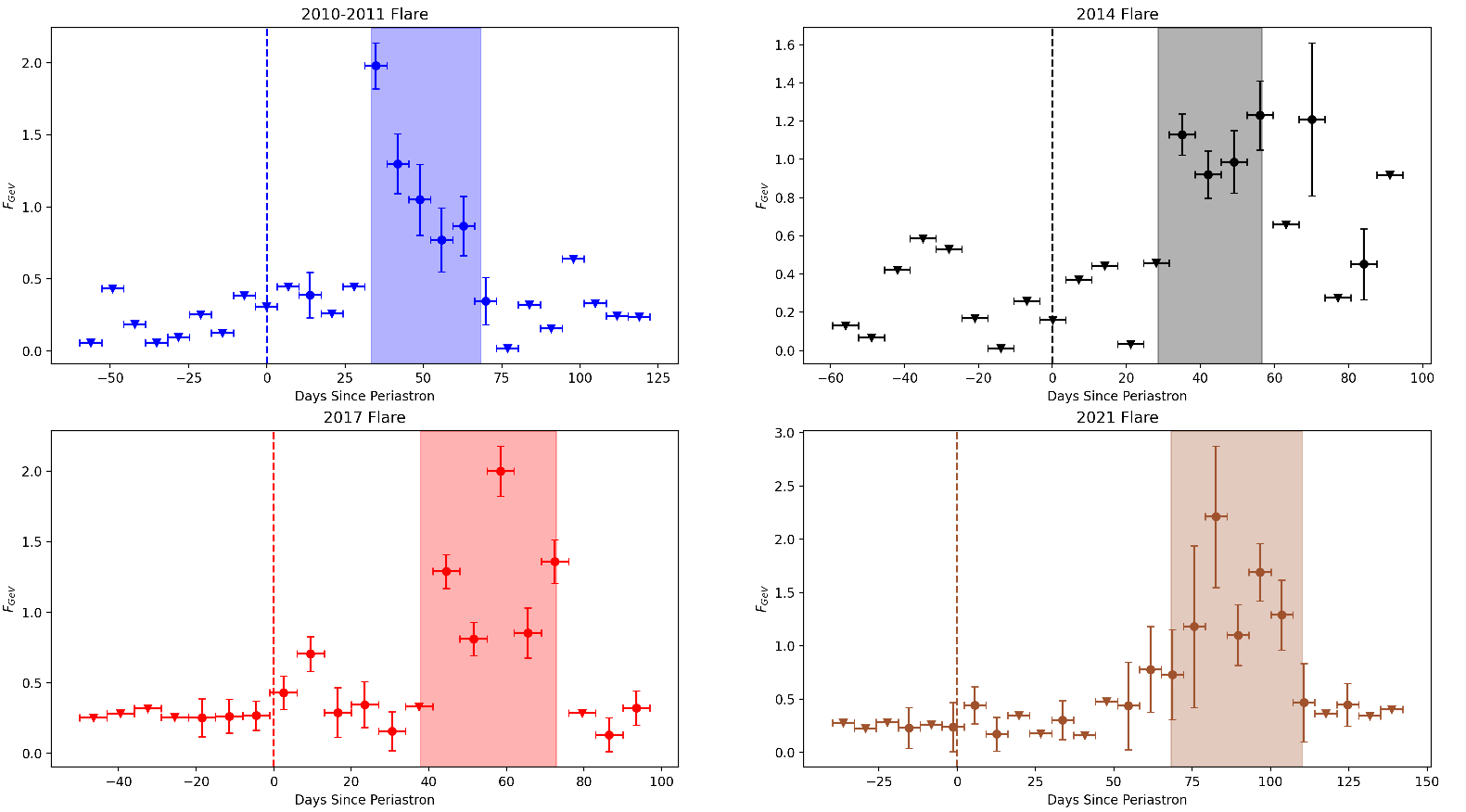}
    \caption{Adapted from Figure 1 in \citealp{psrb1259_2021chernyakova}. $F_{GeV}$ is the \textit{Fermi}-LAT detected photon flux for $E>100$ MeV in units of $10^{-6}$ photons$/cm^{2}/s$ plotted against the number of days since periastron of PSR 1259-63. The dashed vertical line indicates point of periastron and highlighted regions are the consecutive weeks in which FAVA detected a significant long-term flare spatially coincident with this source. } 
    \label{fig:psr_periastron}
\end{figure*}

\subsubsection{PSR B1259$-$63}

We find 21 FAVA flares to be exclusively coincident with the high-mass binary system PSR B1259$-$63, with all flares being part of 4 individual long-term flares.\par

PSR B1259$-$63 is a pulsar that is part of a HMB system, with a massive B2e or Oe star companion (\citealp{Wang_2004_13_yrs_timing_psrb1259_63,Aharonian_2005_psrb1259_63,hess_lat_psrb1259}). This source has a well reported orbital period of $\sim 1237$ days, with an eccentricity of $\epsilon=0.87$ (\citealp{Wang_2004_13_yrs_timing_psrb1259_63}). Prior to, and after periastron, PSR B1259$-$63 traverses the disk-like outflow of its partner star, with the disruption causing emission from radio to TeV energies. See \citealp{2018_Johnson_psrb1259} for a robust description of the system. The crossings were observed by \textit{Fermi}-LAT, with the gamma-ray emission typically peaking $\sim40-60$ days after periastron in 2010 (\citealp{psrb1259_2010abdo,psrb1259_2011tam}), 2014 (\citealp{psrb1259_2015caliandro}) 2017 (\citealp{psrb1259_2018chang,2018_Johnson_psrb1259,psrb1259_2018Tam}), and 2021 (\citealp{psrb1259_2021chernyakova}). The official dates for the previous periastrons of this source are reported as: 2010 December 14 16:39:03 UTC, 2014 May 4 10:02:22 UTC, 2017 September 22 03:25:41, and 2021 February 9 20:48:59 UTC.   \par 


The 21 flares detected by FAVA that are exclusively coincident with this source follow the delayed gamma-ray emission of the disk crossing. The gamma-ray flares detected by FAVA in multi-week epochs are as follows: two 5-week long-term flares (2011 January 17 to 2011 February 21 and 2017 October 30 to 2017 December 4), one  4-week  long-term flare (2014 June 2 to 2014 June 30), and one 7-week long-term flare (2021 April 19 to 2021 May 31). Figure \ref{fig:psr_periastron} is adapted from \citealp{psrb1259_2021chernyakova} which plots the \textit{Fermi}-LAT flux over time. We have added highlighted bands indicating the consecutive weeks of detection by FAVA. Overall, FAVA accurately captures the peak gamma-ray emission as detected by \textit{Fermi}-LAT. \par

\citealp{Prokhorov_2022_newestongaltransients} also reports multi-week flaring behavior of PSR B1259$-$63 in gamma-rays, with a 5 week detection starting on 2017 October 25, and therefore overlapping with one of the long-term FAVA detections of this source.

\subsubsection{LS I +61 303}
LS I +61 303 is coincident with 7 flares exclusively, with no detections in consecutive weeks. \par

LS I +61 303 is another HMB system, with either a black hole or neutron star and a large Be star. One of the first sources whose orbital periodicity was detected in gamma-rays, LS I +61 303 has a $26.6\pm0.5$ day orbital period (\citealp{LSI_2009}). Others propose that LS I +61 303  could be, a magnetar (see \citealp{LSI_magnetar,suvorov2022does}). Nonetheless, the object is known to be transient in gamma-rays. FAVA only detected this source on 7 occasions, and only in the 2009-2019 decade (but not in consecutive weeks).

\subsection{Associated Galactic Sources Detected by FAVA With No Previous Flaring Behavior}\label{sec:gal_assoc}
This section focuses on sources that we report to be flaring in gamma-rays for the first time. In each case, a Galactic gamma-ray source with no known history of gamma-ray flaring behavior was found to be positionally coincident with at least one of the FAVA-detected flares. These sources are PSR J0248$+$6021, PSR J2032+4127, PSR J1731$-$1847, SNR G016.0-00.5, SNR G016.7+00.1, PSR J1826-1256, and SNR G018.1-00.1.
\subsubsection{PSR J0248$+$6021}\label{sec:psrj0148}
Our investigation finds a total of 4 flares coincident with the pulsar PSR J0248$+$6021, 3 of which are solely coincident with this source. Although \citealp{psr_j0248+6021_2010} detected a series of short bursts above the average signal strength of the pulsar, these bursts occurred in radio frequencies and on time periods analogous to the pulsar's spin period of 217 ms.\par
\begin{figure}[h]
    \centering
    \includegraphics[width=0.4\textwidth]{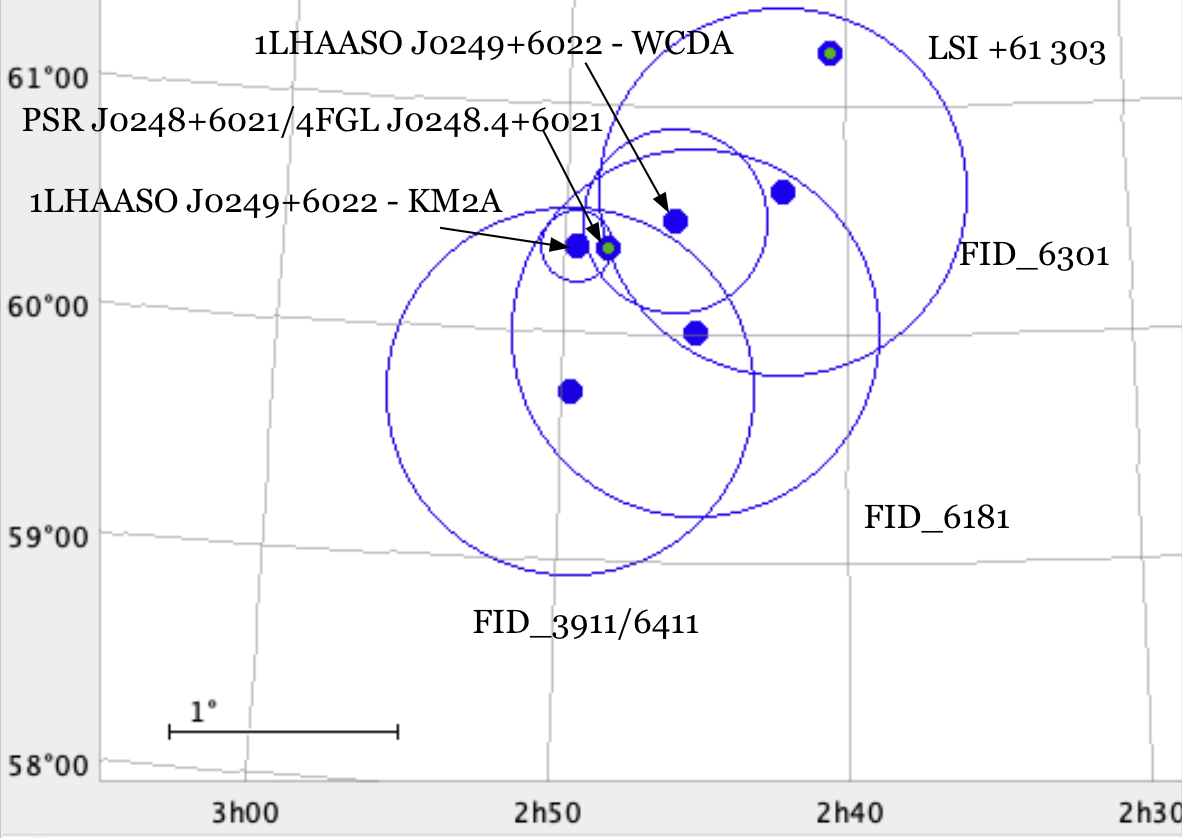}
    \caption{FAVA flare positions and \rad\,  along with LHAASO source positions and statistical \rad\, that are coincident with PSR J0248$+$6021. Nearby 4FGL sources are also labeled.} 
    \label{fig:psr_j0248_fava_lhaaso_4gl}
\end{figure}

The First Large High Altitude Air Shower Observatory (LHAASO) is a TeV air shower detector composed of a Water Cherenkov Detector Array (WCDA) and the Kilometer Squared Array (KM2A) (\citealp{1LHAASO}). The First LHAASO Catalog of Gamma-ray Sources (1LHAASO) reports persistent TeV sources. LHAASO's WCDA data came from 2021 March 05 to 2022 September 30, while  KM2A detections are reported from 2020 January to 2020 to September. Two of the three FAVA-detected flares exclusively coincident with this pulsar occurred in weeks during the time period of the 1LHAASO catalog. Although the 4FGL-DR4 does not report a TeV flag from previous TeV telescopes, both LHAASO detectors report a TeV source coincident with PSR J0248$+$6021. 1LHAASO J0249+6022 is detected by KM2A (RA, DEC: 42.39, 60.37; $R_{95\, stat} = 0.40\degree$, $\Gamma_{KM2A} = 3.82\pm0.18$) with a significance of TS=148.8 and by WCDA (RA, DEC: 41.52, 60.49; $R_{95\, stat} = 0.16\degree$, $\Gamma_{WCDA}=2.52\pm0.16$) at TS=53.3. The position of these sources in comparison to PSR J0248$+$6021 is seen in Figure \ref{fig:psr_j0248_fava_lhaaso_4gl}. We note that $\Gamma_{4fgl}, \Gamma_{LEavg}, \text{and } \Gamma_{HEavg}$ all are in agreement with $\Gamma_{WCDA}$ within $1\sigma$ (see Table \ref{tab:galdr4_src_fav_det}).\par
1LHAASO proposes that the TeV source corresponds to an astrophysical system associated with PSR J0248$+$6021 (e.g a composite SNR). Therefore, we propose that our FAVA detected flares are the first week-long detected gamma-ray outbursts involving PSR J0248$+$6021 and a neighboring system, or nearby feature such as a nebula. 

\subsubsection{PSR J2032+4127}
With two exclusively coincident flares, the pulsar PSR J2032+4127 also provides an interesting potential location of gamma-ray flares. In 2017, the pulsar was identified to be in a binary system with a 15 $M_\odot$ star (\citealp{HO_2017}). With the pulsar displaying X-ray brightening (\citealp{HO_2017}). \citealp{Li_2017_psrxray} conducted long-term X-ray observations and found week-long variability embedded in the long-term increasing trend. \citealp{Li_2018_gev} conducted studies with \textit{Fermi}-LAT but did not detect variability at GeV energies. \par

Due to the vicinity of a nearby TeV source, \citealp{psrj2032_periastron_TeV_2017} used the Major Atmospheric Gamma Imaging Cherenkov telescope (MAGIC; \citealp{MAGIC}) and the Very Energetic Radiation Imaging Telescope Array (VERITAS; \citealp{Veritas}) to observe  PSR J2032+4127 in which they detected gamma-ray emission rising to a factor of 10 above baseline before quickly dropping off a week later. These results led  PSR J2032+4127 to be the second TeV gamma-ray binary system to be detected with the compact object nature known. \citealp{psrj2032_periastron_TeV_2017} model the low state and high state of the outburst as a power law plus a baseline component. The MAGIC detection of the outburst has a best fit of $\Gamma_{MAGIC \, low}=2.57\pm0.26$ and $\Gamma_{MAGIC \, high} = 2.17\pm0.23$, with the low state index agreeing with the LE and HE average indices detected by FAVA (see Table \ref{tab:galdr4_src_fav_det}) \par

These MAGIC/VERITAS observations occurred during periastron (2017 November 13) and do not correspond with the coincident FAVA flares. Six other FAVA flares that are coincident with PSR J2032+4127 are also coincident with Cygnus X-3, so additional analysis is necessary to know if the two flares exclusively coincident with PSR J2032+4127 should actually be attributed to Cygnus X-3. If they are from PSR J2032+4127, these flares would be the first GeV energy flares detected by this source at these week-long timescales. The \textit{Fermi} Light Curve Repository (LCR; \citealp{fermi_LCR}) reports a heightened flare state in weekly time bins for Cygnus X-3 during both flares that are coincident only with  PSR J2032+4127 (FID 6012 and 6911). Therefore, we find it likely that these originate from Cygnus X-3, but further analysis is required to confirm this.

\subsubsection{PSR J1731$-$1847}\label{sec:psrj1731}
PSR J1731$-$1847, a millisecond pulsar (msp), is spatially coincident with a single source. 
This millisecond pulsar is identified to be an eclipsing binary, but this effect would only delay the timed pulses on the order of fractions of a second (\citealp{psr_j1731-1847_2011}). We note that the typical likelihood analysis done by FAVA to better localize this source failed to converge (and therefore giving a \rad \, of $0.8\degree$), and we recommend future analysis in order to identify the origin of this flare, as it could potentially be indicative of the msp interacting with its neighboring environment (i.e. a flaring PWN). If further analysis identifies this to be the origin of the flare, it would be the first gamma-ray flare reported from this  millisecond pulsar's environment.\par






\subsubsection{Flares Coincident with Multiple Galactic Sources Not Known to Flare}
There are two more flares which may have originated from a flaring PWN, but are coincident with more than one non-variable Galactic gamma-ray source. The two flares are:
\begin{enumerate}
    \item FID 3472 - found spatially coincident with SNR G016.0-00.5 and SNR G016.7+00.1
    \item FID 3473 - found spatially coincident with PSR J1826-1256 and SNR G018.1-00.1.
\end{enumerate}
See Table \ref{tab:galdr4_src_fav_det} for additional details on these 4FGL sources and their coincident FAVA flares. We note that both of these FAVA-detected flares were detected in the same week, and within about $1.86\degree$ from one another. FID 3473 was also reported in 2FAV and was labeled as unassociated since the flare could not be firmly established to have originated from PSR J1826-1256 alone. All known Galactic sources that are spatially coincident with these flares are classified as spp, except PSR J1826-1256 (a known pulsar, see Table \ref{tab:galdr4_src_fav_det}). Since the origin of the spp sources is currently ambiguous, we will not discuss them in particular. In order to better understand the nature of these detected flares a robust LAT analysis needs to be applied for the week these flares were detected. However, due to the presence of multiple potential PWN, this environment could potentially host a flaring PWN. \par

\subsection{Potential Crab-Like PWN}\label{sec:potential_flaring_pwn}
Above we discussed the potential association of FAVA flares to pulsars, and possible PWN - both of these environments could potentially mean a flaring PWN is present. Due to the flaring state of Cygnus X-3 during the weeks flares were found coincident with PSR J2032+4127, we find it likely that those flares originate from Cygnus X-3 (which is known to be variable, see Section \ref{sec:cygx-3}). \par
We therefore propose that 4 separate flaring events (FID 3472, 3473, 6181, and 41136), which are positionally coincident with 6 different Galactic gamma-ray sources, could potentially originate from a flaring PWN. None of the coincident sources are reported as variable in the 4FGL-DR4. The Variability Index reported in the 4FGL-DR4 for SNR G016.0-00.5, SNR G016.7+00.1, PSR J1826-1256, SNR G018.1-00.1, PSR J0248+6021, and PSR J1731-1847 is 12.78, 4.00, 8.95, 14.78, 8.14 and 14.66, respectively. Since the 4FGL uses monthly binning for its light curves and this would only be a single week-long outburst at most, we find it unlikely to be detected in the 4FGL analysis. Although the follow-up likelihood analysis that is automatically done by FAVA converged (in most cases) for these flares, the \rad \, of FID 3473 and 3472  are on on the order of $\sim40$ arcminutes ($<0.8\deg)$. Detailed LAT analysis of all of these flares in their respective time windows while implementing improvements to the analysis (e.g. pulsar gating) may enable the identification of the origin of these flares. \par 


Therefore, if any of these known gamma-ray sources are the origin of the flare, this would be the first reported gamma-ray flare reported from any of the sources. More importantly, if this flare originates from  a PWN, that would make this flare the second variable PWN ever detected in gamma-rays. The only other PWN known to be variable in gamma-rays is the one surrounding the Crab nebula (\citealp{rolf_crab_review}).








\subsection{Overview of Unknown/Unassociated FAVA Flare Properties}

Out of the 10 individual unassociated flaring events that survived the cuts we implemented in Section \ref{data}, two of them (FIDs 55913 and 55822) positionally overlap, likely originating from the same source. The distribution of our 10 flares  on the Galactic plane is depicted in an Aitoff projection in Figure \ref{fig:skydist}. 
\par

\begin{figure}
    \centering
    \includegraphics[width=0.5\textwidth]{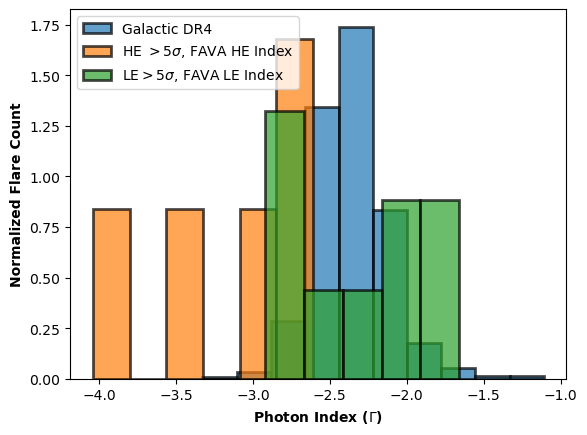}
    \caption{Distribution of power law spectral indices from the likelihood analysis reported by FAVA for our reported unassociated Galactic flaring candidates detected in the HE band (orange) or LE band (green). This is compared with the steady-state average power law fits for identified Galactic sources in the 4FGL-DR4 (blue). Note, most DR4 Galactic sources have their best fit with a \texttt{LogParabola} model but for comparison we plot the fitted power law values. }
    \label{fig:index_hist}
\end{figure}

With a selection criteria of a 5$\sigma$ detection in either the photometric FAVA HE or LE band, we find that 5 unassociated flares are detected exclusively in the LE band (FID 55913, 1281, 6101, 6631, 72921), 1 exclusively in the HE band (FID 57924), and 4 are detected in both bands (FID 55822, 14127, 56019, 60334). See Table \ref{tab:unassoc}. 
The follow-up likelihood analysis implemented in FAVA, with our calibration of its calculated \rad \,, gives a more accurate \rad \, than the photometric localization method alone. Additionally, the maximum likelihood analysis gives an estimation of the photon index of the flare in each band. The smallest \rad\, (including our calibration) is $0.011\degree$ (FID 55822) and the largest is $0.270\degree$ (FID 60334), with a median value of $\sim$0.05$\degree$ across all detections. The mean photon index of all of the LE band detections ($>5\sigma$ for LE) is $-2.31 \pm 0.05$ (of the 5 detected exclusively in this band, the mean is $-2.47 \pm 0.08$). For all the HE-detected flares, the mean photon index in this band is $-3.11 \pm 0.15$ (FID 57924, which is detected exclusively in the HE band, has a HE index of $-2.61 \pm 0.34$). 
The distribution of photon indices of the unassociated flares, compared against the fitted power law photon index ($\Gamma$) of Galactic sources reported in the 4FGL-DR4 are presented in Figure \ref{fig:index_hist}. \par

\subsubsection{Kolmogorov-Smirnov and Anderson-Darling Tests}
In order to help determine if the unassociated Galactic flare candidates likely come from a source population that differs from the known Galactic or blazar sources, we implemented both Kolmogorov-Smirnov (KS) and Anderson-Darling (AD) statistical tests. Using the spectral indices calculated for the flares, we separate the flares into three subcategories: 1) flares spatially coincident with known Galactic sources, 2) flares spatially coincident with known blazar sources, and 3) unassociated flares. Each of these flaring source categories is then split into HE or LE detections. When there is more than one flare spatially coincident with a given source, for each energy band we use the most significantly detected flare in that band. For the unassociated subsets, we only select flares if they have a $>5\sigma$ detection. We then run KS and AD statistical tests on all of the combinations of source subsets and energy selected subsets. \par

Table \ref{tab:KS_AD} outlines the results of these two tests, with the \textit{p}-value results of the AD test above the diagonal of 1's with KS calculated values below. Two major takeaways should be considered. Firstly, when testing the subsets with the LE detected unassociated flares, for both the AD and KS tests, the LE blazar flares have a higher \textit{p}-value than the LE Galactic flares. In either case, the null hypothesis (unassociated sources come from a Galactic, or, blazar population) cannot be rejected. \par

Secondly, when tested with the HE unassociated flares, the HE Galactic flares have a higher \textit{p}-value than the HE blazar flares, and in both tests the HE blazars have \textit{p}-values $<0.02$. This indicates that unassociated flares detected in the HE band are likely not from a blazar parent population.

\begin{deluxetable*}{c| c| c| c |c |c| c }
\centering
   \tablecaption{KS and AD \textit{p}-value Results
    }\label{tab:KS_AD}
    \tablehead{\colhead{}  & \colhead{Gal. LE} &\colhead{Gal. HE} & \colhead{Blz. LE} &  \colhead{Blz. HE} & \colhead{Unassoc. LE}   & \colhead{Unassoc. HE}
    }   
\startdata
\hline
 Gal. LE & 1 & & 0.407 & & 0.5105 &  \\
 \hline
 Gal. HE &  & 1 & & 0.0695 & & 0.1925  \\
 \hline
 Blz. LE & 0.4348 & & 1 & & 0.7025 &   \\
 \hline
 Blz. HE &  & 0.04372 & & 1 & & 0.0125 \\
 \hline
 Unassoc. LE  & 0.5732 & & 0.7904 & & 1 & \\
 \hline
 Unassoc. HE &  & 0.121 & & 0.009243 & & 1  \\
\hline
\enddata
\vspace{0.05cm}
\footnotesize
\textbf{Table \ref{tab:KS_AD}}\\
Values are calculated with  \texttt{ks.test} and \texttt{ad\_test} from R. Values above the diagonal of 1's are the Anderson-Darling calculated \textit{p}-values, while below are the Kolmogorov-Smirnov values. Only matching bands are compared.  

\end{deluxetable*}

\subsection{Potential Counterparts of Unassociated Sources}\label{counterpart}

In an effort to help identify the origin of the unassociated FAVA detected flares reported in Table \ref{tab:unassoc},  we utilized multiple X-ray source catalogs (Living \textit{Swift}-XRT Point Source Catalog (LSXPS, \citealp{LSXPS}), the Fourth XMM-Newton Serendipitous Source Catalog, Thirteenth Data Release (4XMM-DR13, \citealp{4XMM-DR13}), the Chandra Source Catalog  (CSC, \citealp{chandra_ss}), and \textit{Swift}-BAT 105 month Catalog (\citealp{BAT_105m}) as well as analyzed stacked archival data at the flare positions to determine persistent X-ray sources spatially coinciding with the FAVA detected flare. Most of the FAVA detected flares did not have contemporaneous X-ray data. The usage of archival X-ray data implies that we may miss the X-ray counterpart to the flares (if there was one). Therefore, all sources listed in the Appendix should only be taken as possible persistent counterparts.\par

In short, of the unassociated flares, 3 flares are coincident with 4XMM sources (FID 55913, 60334, 72921), 3 flares are coincident with LSXPS sources (FID 57924, 60334, 72921), 1 flare is coincident with a CSC source (FID 14127), and 3 flares are coincident with the \textit{Swift}-BAT 105 month catalog (FID 14127, 55822, 55813). We note that, the coincident BAT detections are already associated to nearby blazars or the Crab Nebula. None of the persistent X-ray sources spatially coincident with one of our Galactic flaring candidates are reported to be transient in their respective catalogs. For a more in-depth description of the X-ray catalogs and the fields of each unassociated FAVA detected Galactic candidate, please see the Appendix.

\subsubsection{Flares that lack X-ray Data}
Beyond the X-ray catalogs, the High Energy Astrophysics Science Archive Research Center (HEASARC; \citealp{heasarc}) was also searched for possible X-ray data (from \textit{Swift}-XRT, \textit{XMM}, \textit{Chandra}, and \textit{NuSTAR}) at the location of each of these FAVA detected flares. Three of the flares detected by FAVA have no archival X-ray data from the above mentioned missions. These flares are FID 
1281, 6631, 
and 6101. 
\begin{longrotatetable}
\begin{deluxetable*}{c c c c c c c c c c | c c c }
   \tablecaption{Unassociated FAVA Detected Galactic Flare Candidates$^a$
    \label{tab:unassoc}}
    \tablehead{\colhead{Flare ID} &  \colhead{RA} & \colhead{DEC } & \colhead{\rad \,$^{C}$} & \colhead{LE $\sigma$$^b$}  & \colhead{HE $\sigma$$^c$} & \colhead{LE $\Gamma$$^d$} & \colhead{HE $\Gamma^e$}  & \colhead{Start Date} & \colhead{End Date} &\colhead{Closest} & \colhead{4FGL$^g$} & \colhead{Dist/\rad$^h$} \\
    \colhead{} &  \colhead{($\degree$)} & \colhead{($\degree$)} & \colhead{($\degree$)} & \colhead{} & \colhead{} & \colhead{} &  \colhead{} & \colhead{M/D/Y} & \colhead{M/D/Y}& \colhead{ 4FGL$^f$} & \colhead{Class}& \colhead{} 
    }   
\startdata
& & & & & & &   Multiple$^i$ & & & &  \\
\hline
\hline
\hline
\hline
\hline
55822 & 278.46 & -21.05 & 0.011 & 40.0$^*$ & 40.0$^*$ & -2.12 $\pm$ 0.03 & -2.93 $\pm$ 0.08   & 04/08/19 & 04/15/19&  J1833.6$-$2103 &FSRQ&  4.1 \\
55913$^\bigcirc$ & 278.46 & -21.05 & 0.024 & 25.8 & 3.7 & $-2.25\pm 0.04$ & $-2.29 \pm 0.87$  & 04/15/19 & 04/22/19 &  J1833.6$-$2103 &FSRQ& 1.7 \\
\hline
\hline
 & & & & & & &  Single$^j$ & & & &  \\
\hline
\hline
1281 & 194.81 & -65.46 & 0.058 & 7.5 & 1.5 & -2.73$\pm$0.25 & -1.03$\pm$0.48  & 01/10/11 & 01/17/11 & J1302.9$-$6349 & HMB & 28.6\\
6101 & 138.01 & -56.99 & 0.040 &12.2 & 3.7 & -1.66$\pm$0.06 & -2.98$\pm$0.19  & 04/06/20 & 04/13/20 &  J0904.9$-$5734 & bcu &  28.2 \\
6631 & 196.71 & -64.75 & 0.169 &  10.2&0.6 & -2.79$\pm$0.16 & -0.81$\pm$0.48  & 04/12/21 & 04/19/21 & J1302.9$-$6349 & HMB &  6.0\\
14127 & 83.69 & 22.01 & 0.014 & 40$^*$ & 10.3 & -2.06$\pm$0.03 & -4.04$\pm$0.29 & 04/11/11 & 04/18/11  & J0534.5+2200 & Crab Pulsar  &  4.1\\
56019 & 278.49 & -21.1 & 0.019 & 40$^*$ & 40$^*$ & -1.81$\pm$0.03 & -2.63$\pm$0.07  & 04/22/19& 04/29/19 &  J1833.6$-$2103  & FSRQ & 4.1\\
57924 & 135.95 & -57.61 & 0.104 &  3.0 & 5.5 &$-2.52\pm 0.38$&  $-2.61 \pm 0.34$& 09/02/19 & 09/09/19  &  J0904.9$-$5734 & bcu & 1.42\\
60334 & 278.25 & -21.36 & 0.270 & 10.7 & 7.4 & $-2.44 \pm 0.2$ & $-3.36\pm 0.57$ &2/17/2020 & 2/24/2020 & J1833.6-2103 & FSRQ & 1.24\\
72921 & 278.43 & -20.81 & 0.077 & 5.6 & 0.3 &  $-2.92\pm0.29$  & $-1.99\pm0.49$   & 07/18/22 & 07/25/22 &  J1833.6$-$2103  & FSRQ & 3.3\\
\hline
\enddata
\vspace{0.05cm}
\footnotesize
\textbf{Table \ref{tab:unassoc}} \\
$^a$ All FAVA reported associations for these detected flares listed `None'\\
$^{C}$ Calibrated \rad \, (i.e. scaled by 1.06. See Section \ref{calibration}) \\
$^b$  Significance of detection for $E = 100-800$ MeV. The maximum possible value is 40$\sigma$ \\
$^c$  Significance of detection for $E = 800-300000$ MeV. The maximum possible value is 40$\sigma$\\
$^d$  Fitted power-law index for  $E = 100-800$ MeV \\
$^e$  Fitted power-law index for  $E = 800-300000$ MeV\\
$^f$  Source in the 4FGL closest to the FAVA detected flare by angular separation. All sources listed officially have the identifier `4FGL' prior to the listed name. All sources listed have a Variability Index $>27.69$ as reported by the 4FGL-DR4 \\ 
$^g$ Class designations listed are reported from the 4FGL-DR3
$^h$ Ratio of the angular separation distance between the nearest 4FGL source and the \rad \,\, as reported by FAVA's follow-up localization, rounded to the nearest tenth. \\
$^i$ For clarity, the detections that likely originate from the same source are grouped together \\
$^j$ All flares listed in this section were not coincident with any other \\
$^*$ Is hitting the maximum significance threshold as reported by FAVA, and so could be higher. \\
$^\bigcirc$ - Ratio distance $<4$ and  \textit{Fermi} Light Curve Repository reports a heightened state of the nearby 4FGL source 
\end{deluxetable*}
\end{longrotatetable}

\section{Discussion} \label{discussion}
Due to the bright Galactic diffuse emission, as well as the systematic uncertainty, \textit{Fermi}-LAT is biased against detecting flaring sources in the Galactic plane. In this work we used 14.98 years of data (779 weeks) to study the population of Galactic gamma-ray transients. 

The 4FGL-DR4 only reports firm associations for 32 sources classified as gamma-ray binaries, novae, and PWN. With so few known Galactic sources detected in gamma-rays (we exclude millisecond pulsars in our discussion since the transient nature caused by their fast rotation is well studied), and even fewer of these source classes exhibit transient behavior (9 as reported by \citealp{2FAV}), this investigation uses \textit{Fermi}-LAT All-sky Variability Analysis (FAVA) to identify potential Galactic candidates that would flare on week-long time periods. \par

After FAVA identifies flares with its photometric method, it also does an additional likelihood analysis of the detected flares,  reporting spectral properties along with a better localized position ($<0.8\degree$). For the unassociated flares that we posit to be Galactic flare candidates (see Table \ref{tab:unassoc}), of the flares detected exclusively in the HE band, the mean of the HE photon index is $\Gamma_{HEavg} = -2.40 \pm 0.28$ and of those detected exclusively in the LE band, the average LE photon index is $\Gamma_{LEavg}=-2.47 \pm 0.19$. Of the 4 flares detected with $>5\sigma$ significance in both bands, their average HE index is $\Gamma_{both\, HE} = -3.24 \pm 0.16$ and their average LE index is $\Gamma_{both\, LE} = -2.11 \pm 0.05$. This seems to indicate that for these unassociated flares, when significantly detected in both bands they are generally harder in the LE band and softer in the HE band compared to the flares detected in a single energy band. Comparatively, the average photon index for a Galactic source reported in the 4FGL-DR4 is $\Gamma_{Galactic\, 4FGL} = -2.33\pm 0.01$, with PWN having the hardest average, ($\Gamma_{PWN} = -2.08 \pm 0.02$) and novae having the softest ($\Gamma_{NOV} = -2.46\pm 0.06$). \par

We draw specific attention to FID 
1281, 6631, 
and 6101 as each of these flares was significantly detected and occurred at a location in the sky with absolutely no archival X-ray observations. We highly recommend that these regions be systematically searched in follow-up campaigns to identify possible counterpart candidates to the origin of the FAVA detected flare. Although possibly coincidental, we note that each of these flares were all detected by FAVA in the LE and not the HE band (at $\ge5\sigma$).\par



\subsection{Estimating Flare Rates of Galactic Origin}\label{flare_estimates}

We find 115 flares to be coincident with at least 1 known Galactic gamma-ray source. Additionally we filtered the unassociated flares (as detailed in Section \ref{pipeline_filter}) to remove blazars. Our filter for selecting likely Galactic candidates, is able to correctly vet blazars from our data set with an accuracy of 68.4\% (based on the use of the WISE blazar strip alone). Our testing of the WISE blazar strip gave us a false negative rate (FNR = 100 - TPR) of 24.3\% and a false positive rate (FPR = 100 - TNR) of 39.3\%. Filtering flares based on sources in their \rad\, with the WISE blazar strip resulted in 10 individual flares remaining. Based on positional coincidence, we posit that these 10 flares originate from 9 distinct flaring Galactic candidates.\par 


The false negative rate allows us to estimate that 2-3 of our 10 unassociated flares are potentially produced by blazars. In tandem with our estimates of possible wrongful identifications (see Section \ref{sec:misplaced}) we estimate the detection of 9 unassociated likely Galactic gamma-ray transients. Moreover, our false positive rate means that we likely filter out 3$-$4 additional flaring Galactic candidates that were detected by FAVA. \par
We then estimate that over the course of the 779 weeks studied, 12-13 unassociated Galactic flares occurred. Combining these unassociated flares with the 105 FAVA flares coincident with only a single Galactic gamma-ray source as reported in the 4FGL-DR4,  we estimate a total of 117-118 flaring events that are likely of Galactic origin during our time of study. If we count FAVA flares that coincide with more than one known Galactic gamma-ray source, this estimate becomes 127-128. Employing the estimate from all coincident FAVA and 4FGL-DR4 Galactic sources, along with likely Galactic candidates from this work, we estimate that there are $\sim8.5$ Galactic gamma-ray transient events per year. Of those 8.5 flares, this exploration has led to the estimate that each year, $\sim 1$ of those comes from an unassociated flare, $\sim 1$ originates from a classical nova, $\sim 2$ are from Cygnus X$-3$, $\sim 2$ stem from the Crab Nebula, with the other $\sim2.5$ coming from other Galactic gamma-ray sources.

\subsection{Likely Misidentified Detections and Most Likely Sources of Galactic Origin} \label{sec:misplaced}
Some unassociated flares fall in the vicinity of nearby 4FGL sources (typically blazars). We want to establish whether, statistically, a flare would likely be caused by a nearby variable source. We establish a `ratio distance' by taking the ratio of the distance from the flare to the nearest 4FGL source with the FAVA reported \rad\, (i.e. dist/\rad). In essence, this establishes how many $\sigma$ apart the two locations are - as in, how much disagreement their positions have. If the $\sigma$ is large, it means that the flare is unlikely to have originated from the nearby source. On the other hand, if $\sigma$ is small, it means that the flare potentially could have originated from that nearby source.  For the purposes of identifying the most promising flares,  we establish a ratio distance of $\geq4$ as the cutoff for flares that are likely not a misidentified detection of the nearby source (since their distance apart compared to their localization is large). That is, the nearby 4FGL source has a $\sim$0.006\% chance of being of the same origin as the detected flare based on statistical properties alone. \par

Furthermore, if the nearby source has a  a Variability Index  $>27.69$ as reported in the 4FGL, it has $<1\%$ chance of being a steady source (\citealp{4fgl-DR4,Scargle_2013}). Four of our FAVA detected Galactic flare candidates reported in Table \ref{tab:unassoc} are at a ratio distance of $<4$. Each of these four detected flares, the closest 4FGL source is likely variable. We suspect that these may be misidentified as new detections and are likely a flare caused by the nearby source.\par



Moreover, we checked these four unassociated FAVA Galactic flare candidates suspected of having originated from a their nearby 4FGL source by checking the closest 4FGL source in the \textit{Fermi} Light Curve Repository (LCR; \citealp{fermi_LCR}) using weekly time bins. FSRQ PKS 1830-211 (the closest 4FGL source to FID 55913) was flaring from a low to high state in one of its strongest flares recorded during the same week as FID 55913. For FID 57924, the nearest 4FGL source is PKS 0903-57 (a bcu) and during the week of the FAVA detected flare this blazar is slightly above the baseline variation, but not in a distinctively flaring state. PKS 1830-211, which is also the nearest 4FGL source for FID 60334, according to the LCR this source is actually in a decaying state from a large flare that occurred in late December of 2019. Lastly, PKS 1830-211 (also the nearest source for FID 72921) is at a relatively stable state during the week of FID 72921's detected flare. Overall this leads us to believe that FID 55913 is most likely a flare originating from the blazar, but the other three (FID 57924, 60334, 72921) could still potentially be from an unknown source. Therefore, we warn readers that FID 55913 is likely a flare from a blazar.

\section{Summary} \label{conclusion}
Outside of the large number of pulsars detected at high-energies, only three other Galactic source classes are known to emit gamma-rays in a transient manner: gamma-ray binaries, novae, and the Crab Nebula Pulsar Wind Nebula. Of these source classes, the 4FGL-DR4 only reports persistent gamma-ray emission from 32 firmly identified sources (60 associated sources). Moreover, low statistics of Galactic gamma-ray transients for each of these source classes means the discovery of each additional Galactic gamma-ray transient adds substantially to the statistics of the source class it belongs to. 

As part of utilizing FAVA to identify new Galactic gamma-ray transients, we have been able to:
\begin{itemize}
    
    \item Estimate a yearly Galactic gamma-ray transient rate of $\sim8.5$ flares/year. The composition of this yearly rate is $\sim2$ flares from the Crab Nebula, $\sim2$ flares from Cygnus X-3, $\sim1$ flare from a classical nova, $\sim1$ unassociated source, with the remaining $\sim 2.5$ flares coming from other known Galactic gamma-ray sources. 

    \item Report the first gamma-ray flare spatially coincident with 6 4FGL sources with no prior gamma-ray outbursts. These are: PSR J0248+6021, PSR J1731-1847, SNR G016.0-00.5, SNR G016.7+00.1, PSR J1826-1256, and  SNR G018.1-00.1. SNR G016.0-00.5, SNR G016.7+00.1, and SNR G018.1-00.1 are all classified as either a supernova remnant, or a pulsar wind nebula. PSR J0248+6021, PSR J1731-1847, and PSR J1826-1256 are all pulsars. If follow-up analysis confirms that the flare does originate from one of these source environments, it would potentially mean the confirmation of a flaring pulsar wind nebula (PWN), and possibly the discovery of the second gamma-ray variable PWN like the Crab Nebula's PWN.  
    
    \item Determine 10 individual unassociated flares that are of likely Galactic origin. 
    
    \item Use the WISE blazar strip for filtering out possible blazars based on infrared counterparts in their gamma-ray \rad. Furthermore, we tested the WISE blazar strip using the gamma-ray locations and \rad\, of sources reported in the 4FGL. We report the WISE blazar strip to have a 68\% accuracy rate for correctly distinguishing blazars and Galactic sources given their gamma-ray detections. We also verify that classical novae with gamma-ray emission are unlikely to be filtered out (94\% are not identified as blazars). 


    \item Report 18 long-term flares ($>$1 week) from consecutive, coincident FAVA detections. These 18 long-term flares likely come from 4 unique Galactic gamma-ray sources. We expect at least 1 long-term flare per year to be detected by FAVA. 

    \item Potential observational confirmation of the Crab Nebula's PWN `wave' outburst behavior predicted by \citealp{crab_striani_2011}.  
    
    \item Identify spectral parameters for each unassociated flare. We find that, on average, when compared to other unassociated flares detected in a single band, unassociated flares detected significantly in both LE and HE bands tend to be harder in the LE and softer in the HE. The mean photon index for all the unassociated flares detected in the LE band is $-2.34\pm0.23$ and for all unassociated flares detected in the HE band is $-2.44\pm0.43$.
    
    \item Report (in the Appendix) persistent X-ray sources from archival data that are spatially consistent with the FAVA detected flares. The potential Galactic flaring candidates with a persistent X-ray source coincident with their \rad\, are FID 14127, 55822, 55813, 55913, 57924, 60334, and 72921. 
    
    \item Recommend fields for future X-ray study since they had no X-ray observations where a likely flare of Galactic origin was detected. These flares are FIDs 1281, 6631, and 6101.

\end{itemize}

Future follow-ups and analysis of the fields with FAVA detected gamma-ray flares of likely Galactic origin could lead to the detection of new transient sources. Most importantly, it could verify a potentially new variable PWN. \par

\software{
    Astropy (\citealp{Astropy}),
    numpy (\citealp{Numpy}),
    R (\citealp{R_package}),
    SAO Image DS9 (\citealp{DS9}),
    TOPCAT (\citealp{topcat}),
    XSPEC (v12.11.1; \citealp{Arnaud_Xspec})
}
\newline \\


We decorously thank NASA for the generous support in funding the \textit{Fermi} proposal "Studying the origin of historical Galactic transients with FAVA" (Proposal ID: 161091) which funded this research. We graciously thank Dr. Chernyakova of Dublin City University and Dr. Dmitry Prokhorov of the University of Amsterdam for their openness to share their data with regards to the flares they studied as well as Dr. Shaoqiang Xi who helped with accessing the 1LHAASO catalog.\par This research has used the SIMBAD database, operated at CDS, Strasbourg, France. This research has made use of data and/or software provided by the High Energy Astrophysics Science Archive Research Center (HEASARC), which is a service of the Astrophysics Science Division at NASA/GSFC and the High Energy Astrophysics Division of the Smithsonian Astrophysical Observatory. This publication made use of data products from the Wide-field Infrared Survey Explorer, which is a joint project of the University of California, Los Angeles, and the Jet Propulsion Laboratory/California Institute of Technology, funded by the National Aeronautics and Space Administration. Lastly, we thank the reader for his or her interest in our work.


\appendix \label{Appendix}

\section{Codes}
\begin{itemize}
    \item Code used to webscrape data from the FAVA website: \url{https://github.com/dankocevski/pyFAVA}
    \item 3D WISE blazar strip checking  AllWISE sources in \rad \,with upper limits included: \url{https://github.com/tryingastronomy/Blazar_codes/tree/main/AW_blazar_strip} 
\end{itemize}

\section{X-ray Catalog Descriptions}
We cross-matched our 10 unassociated flares using the best \rad\, from the maximum likelihood analysis of the FAVA detected flare with various X-ray catalogs. The catalogs are described below. 

\subsection{Swift-BAT 105 Month Catalog} \label{SwiftBAT105}
Using $2.6'$ for \textit{Swift}-BAT's \rad\footnote{see \url{https://heasarc.gsfc.nasa.gov/docs/heasarc/caldb/swift/docs/bat/SWIFT-BAT-CALDB-CENTROID-v2.pdf}}, we find 3 flaring Galactic candidates coincident  with the \textit{Swift}-BAT 105 month catalog. These are SWIFT J0534.6+2204 (Crab nebula) with flare 14127, SWIFT J1833.7-2105 (PKS 1830-21) with flares 55822 and 55913. These cross-matches, and the other possible counterparts for each FAVA detected Galactic flaring candidate are discussed in detail in Section \ref{likely_ctpt}. \par

\subsection{LSXPS}
The Living \textit{Swift}-XRT Point Source (LSXPS; \citealp{LSXPS}) catalog updates its collection of point sources detected by \textit{Swift}-XRT in real time. LSXPS covers 5,371 deg$^2$, and when used for this manuscript was updated through 2023 October 25.
LSXPS includes a method for detecting transients. This is done by 1) determining if it is a cataloged X-ray source, and 2) compares the measured flux to the source's historic upper limits. See Section 4 of \citealp{LSXPS} for more details. Cross-matching our FAVA detected flares with the LSXPS transient catalog found no matches that were spatially coincident with one another. We note that a lack of X-ray monitoring over the fields of interest rather than the non-transient nature of the X-ray sources may be the cause of no cross-matches being detected.

\subsection{4XMM-DR13}
The Fourth XMM-Newton Serendipitous Source Catalog, Thirteenth Data Release (4XMM-DR13; \citealp{4XMM-DR13}), released 2023 June 12, includes  13,243 XMM-Newton EPIC observations from  2000 February 3 $-$ 2022 December 31, covering over 1328 deg$^2$ of the sky. The 4XMM-DR13 also includes a \texttt{SC\_VAR\_FLAG} for the most variable detection of a source, \texttt{SC\_Fvar} which reports the lowest probability that the source is constant, and \texttt{SC\_chi2prob} which is the $\chi^2$ probability that the unique source detected by one of the previous observations is constant. Out of the 54 sources that are spatially compatible with an unassociated FAVA flare, none were flagged as variable.  




\subsection{CSC 2.0} \label{csc_section}
The \textit{Chandra} Source Catalog 2.0 (CSC; \citealp{chandra_ss,chandra_ss2}) includes 317,167 unique compact and extended X-ray sources detected by \textit{Chandra} up until December 31, 2014. One of the 10 unassociated flaring Galactic source candidates are coincident with at least 1 \textit{Chandra} detection (FID 14127).


\section{Possible Counterparts and X-ray Analysis in the Field of FAVA Detected Galactic Flares} \label{likely_ctpt}
For the 10 Galactic flaring candidates, 7 have archival X-ray data taken near the  location of the detected event. Below, we outline potential counterpart candidates to the FAVA detected Galactic flaring candidates, starting with what we find in their X-ray fields. For those lacking X-ray data, we also discuss the known sources within each flare's \rad \,\, that are reported by SIMBAD (\citealp{Simbad}). We recommend X-ray surveys of all areas of the sky covered by our flares, especially the fields that lack any X-ray observation. It is very possible that the individual counterpart responsible for the gamma-ray flare detected by FAVA is not detectable with the limited archival X-ray data we used. Since the X-ray observations do not coincide temporally with the FAVA detected flare, these reported counterpart candidates are simply the best counterpart candidates we have available at this time.  




\begin{sidewaysfigure*}
\centering
\vspace{+3cm}
\hspace{+2.5cm}
\includegraphics[width=1.1\textwidth]{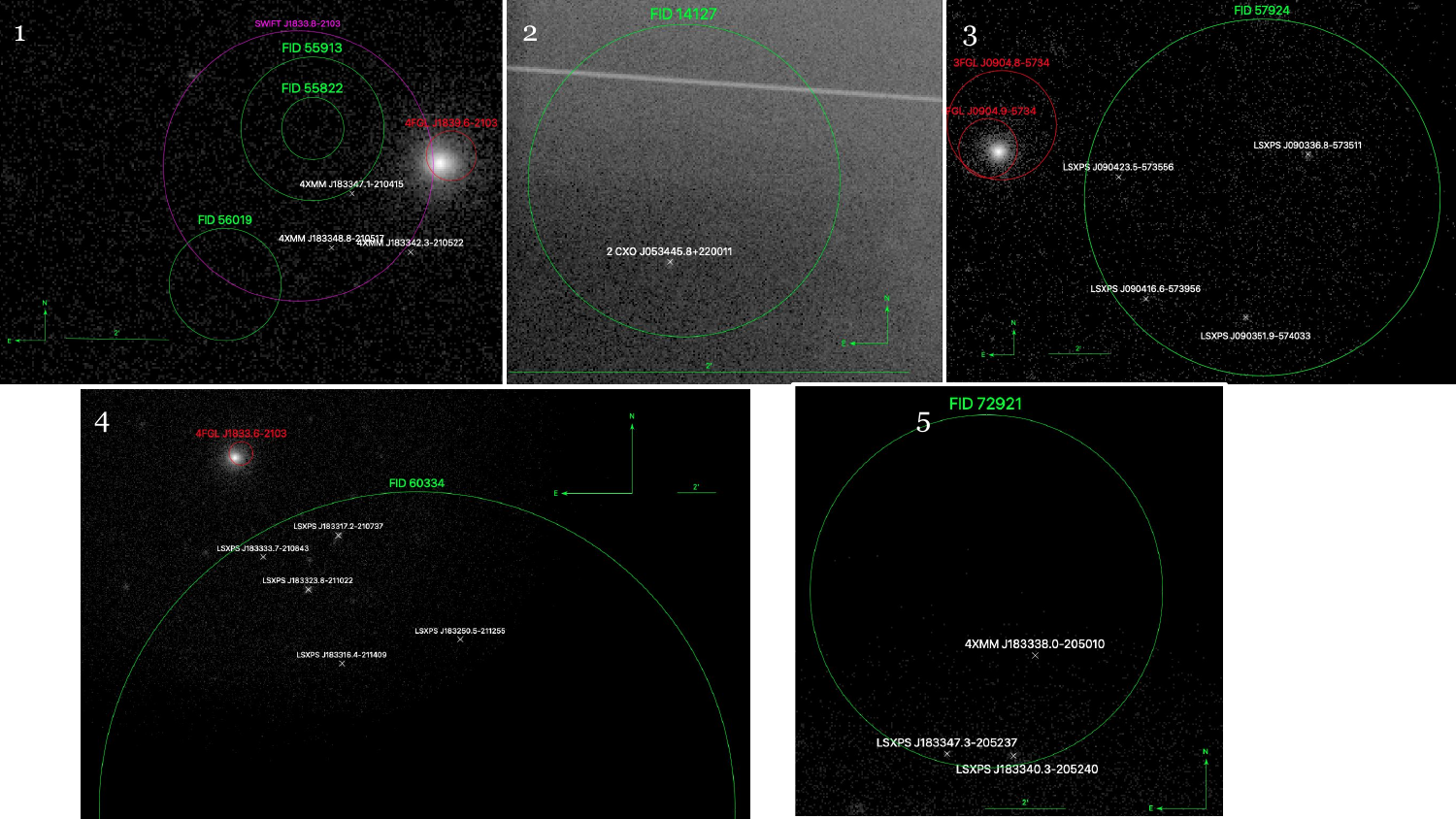}
\caption{FAVA detected flares with X-ray data in the same FOV are plotted here with their corresponding FAVA detection (green), nearby 4FGL sources (red), 4XMM and LSXPS survey sources (white), and \textit{Swift}-BAT coincident sources (magenta). All fields were observed using \textit{Swift}-XRT (with the exception of Panel 2 which was observed with \textit{Chandra}). FIDs 1281,6101, and 6631 lack X-ray data. FID 7401 is not included since it corresponds to GRB 221009A .  }
\label{fig:panel_images1}
\end{sidewaysfigure*}

\subsection{Flare ID: 55913, 55822} \label{fdgfc2}
\color{black}
Flare 55913 and flare 55922 are coincident with one another. The nearest 4FGL object is the FSRQ, 4FGL J1833.6$-$2103. This variable object has been associated with the \textit{Swift}-BAT detection, SWIFT J1833.8$-$2103 which is coincident with both flares. The ratio distance (distance to 4FGL object divided by FAVA detected \rad) is 1.7$\sigma$ and 4.1 $\sigma$ for the different flares. As discussed above, FID 55913 likely is a misidentified flare that likely originates from this blazar. 


\subsection{Flare ID 1281}
Flare 1281 is a ratio distance of 28.6$\sigma$ away from the high mass binary (HMB), 4FGL J1302.9-6379, the nearest 4FGL source, which can be completely excluded as the counterpart based on distance alone. No archival X-ray data is present for this field. SIMBAD reports a few sources coincident with the FAVA detected flare, including a classical Cepheid variable, a spectroscopic binary, a RR Lyrae Variable, and a couple Long-Period Variable stars. None of these sources are  convincing counterpart candidates to be the origin of a 100$-$800 MeV flare. We recommend future studies of this region of the sky to identify the origin of this 7.5$\sigma$ LE flare.  


\subsection{Flare ID 6101}
\color{black}
The nearest 4FGL source is the blazar, 4FGL J0904.9-5734, at a ratio distance of 28.2$\sigma$. This large of a ratio means that this source can be excluded as the source of the flare. Flare 6101 also lacks X-ray observations in its \rad\,. Searching in SIMBAD only gives stars coincident with the position of the flare. Detected with a $12.2\sigma$ significance in the LE band and $3.7\sigma$ significance in the HE band, follow-up observations are necessary to properly study the origin of this flare. 

\subsection{Flare ID 6631}
In the neighborhood of flare 1281, FID 6631's detected flare's nearest known 4FGL neighbor is also 4FGL J1302.9-6379, with a ratio distance of 6.0$\sigma$. No X-ray data from \textit{Swift}-XRT, \textit{XMM}, or \textit{Chandra} is available in the region of flare 6631. SIMBAD reports a multitude of stars of various types, but no object that would be a convincing counterpart to this significantly detected flare. Follow-up surveys of this region are required to discuss the possibilities of the origin of this flare more in depth.

\subsection{ Flare ID 14127 }\label{FID: 14127}

Detected on the outskirts of the Crab Nebula, flare 14127 is a ratio distance of 4.13$\sigma$ to the pulsar at the center of the Crab's position. The ratio distance to the nearest point in the Crab's nebula (as detected in X-rays) is 2.24$\sigma$. The FAVA detected flare occurs well outside of the Crab Nebula itself. Checking the \textit{Fermi} LCR, (which only reports the nebula's synchrotron component) the Crab's PWN has an increase in its flux between 2011 April 8 - 15, going from 7.97$\pm 1.78\times 10^{-7}$ to $1.74\pm0.17 \times 10^{-6}$ photons/cm$^2$/s in the 0.1-100 GeV band. The average photon flux for the Crab PWN as reported through December 1, 2023 from the LCR is 1.26$\times 10^{-6} \pm 1.1\times 10^{-11}$ photons/cm$^2$/s . Although this week is above the mean value, it is not a dramatic departure. This, paired with the distance to the edge of the Crab points to this being a potential detection of a new source.  \par 

Flare 14127 has a \rad \,\, that has minimal overlap with the uncertainty area of the \textit{Swift}-BAT detection of the nebula (SWIFT J0534.6+2204). Within the FAVA detected flare's \rad \,\,, there is a CSC source present.  2CXO J053445.8+220011 was detected on January 27, 2004 at 5.87$\sigma$.(\citealp{chandra_ss,chandra_ss2}). 
 Although detected in 2004, no other observations of this part of the sky have been observed. Therefore, at this time the Chandra detected X-ray source appears to be the most likely candidate for the origin of the FAVA detected flare. We recommend further studies of this X-ray source, and this field to solidly confirm the origin of this flare.

\subsection{ Flare ID 56019}
\color{black}
Occurring in the same region of the sky as FIDs 55913, and 55822 FID 56019 was detected in both the HE and LE bands at a significance of $40\sigma$ (the highest value that FAVA reports). With a ratio distance of 4.1$\sigma$ from the nearby source, 4FGL J1839.6-2103, FID 56019 likely originates from a uniquely different source than FIDs 55913 and 55822.  No persistent X-ray sources were spatially coincident with this detection 

\subsection{Flare ID: 57924}
The \rad \,\, of flare 57924 does not overlap with, but is nearby 4FGL J0904.9$-$5734, the highly variable (Variability Index = 8488.27) BL Lac object, although it was not flaring during the week of this FAVA detected flare. The ratios of distance to the nearest 4FGL source divided by the FAVA \rad \,\, is 1.42. This flare has 4 LSXPS sources spatially coincident with its position.


Searching HEASARC, this flare has both archival \textit{Swift}-XRT and \textit{Chandra} data in the neighborhood of the flare. For the \textit{Chandra} observation, we find the detected flares to be outside of the field of view (FOV). 
The most significant detection from the LSXPS J090336.8$-$573511 which is coincident  with flare 57924 has a signal-to-noise ratio (SNR) of SNR=5.5. LSXPS J090351.9-574033 was detected at SNR = 3.9. 

\subsection{Flare ID 60334}
\color{black}
With the largest \rad\, out of the unassociated flares, the distance to the nearest 4FGL source divided by the \rad\, is only 1.24. The nearest 4FGL source is the FSRQ, 4FGL J1833.6$-$2103. This 4FGL source has a variability of 37330 reported in the DR4, but is not varying during the week of the flare. Only part of the \rad\, of FID 60334 has X-ray data with a number of LSXPS point sources. Since only a small portion of the \rad\, area has been observed it is hard to determine if there is a proper persistent X-ray counterpart. This flare has 41 spatially coincident 4XMM sources and 15 LSXPS sources.

\subsection{Flare ID 72921}
The closest detected 4FGL source to this flare is 4FGL J1833.6$-$2103 at a ratio distance of 3.3. Falling on the outskirts of the \textit{Swift}-XRT FOV, only a small portion of the FAVA \rad \,\, has counts. Even in this small area of the \rad, multiple sources from LSXPS and 4XMM-DR12 are present. Two of the detections overlap and coincide with the same source, those being LSXPS J183340.3$-$205240 and 4XMM J183340.2$-$205241. Due to the lack of observation of the entire FAVA detection, until additional observations are taken of this region, these persistent X-ray sources are our only counterpart candidates. This source has 4 spatially coincident 4XMM sources, and 2 from LSXPS.

\newpage
\bibliography{bibliography}
\bibliographystyle{aasjournal}

\end{document}